\documentclass[10pt]{iopart}
\usepackage{epsfig,bm}
\usepackage{graphicx}
\expandafter\let\csname equation*\endcsname\relax
\expandafter\let\csname endequation*\endcsname\relax
\usepackage{amsmath}
\usepackage{placeins}
\usepackage{amssymb,psfrag,textcomp}
\usepackage{amsthm,upgreek}
\DeclareMathAlphabet{\mathdutchcal}{U}{dutchcal}{m}{n}
\SetMathAlphabet{\mathdutchcal}{bold}{U}{dutchcal}{b}{n}
\DeclareMathAlphabet{\mathdutchbcal}{U}{dutchcal}{b}{n}

\usepackage[usenames,dvipsnames,svgnames,table]{xcolor}

\usepackage[
  backend=biber,
  style=phys,
  sorting=none,
  biblabel=brackets,   
  doi=false,           
  url=false,           
  isbn=false,
  eprint=false
]{biblatex}
\usepackage{hyperref}
\hypersetup{colorlinks=true, linkcolor=NavyBlue, citecolor=PineGreen,urlcolor=cyan}

\newcommand{\be}{\begin{equation}}
\newcommand{\ee}{\end{equation}}
\newcommand{\bd}{\begin{displaymath}}
\newcommand{\ed}{\end{displaymath}}

\newcommand{\re}{\mathrm{Re}}
\newcommand{\im}{\mathrm{Im}}

\newcommand{\mathxy}[1]{\tilde{\mathrm{#1}}}

\newcommand{\citetemp}[1]{[\textbf{pending}]}
\newcommand{\exValueF}[1]{\langle #1 \rangle_{\bm F}}
\newcommand{\exValuebF}[1]{\left\langle #1 \right\rangle_{\!\!\bm F}}
\newcommand{\exValueb}[1]{\left\langle #1 \right\rangle}
\newcommand{\Smat}{\bm{S}}

\begin{document}

\title{Spectral properties of a Non-Hermitian extension of the diluted Wishart ensemble}
\author{Edgar Guzm\'an-Gonz\'alez}
\ead{edgar.guzman@hainanu.edu.cn}
\address{School of Physics and Optoelectronic
Engineering, Hainan University, 570228 Haikou, P. R. China}

\author{Isaac P\'erez Castillo}
\ead{iperez@izt.uam.mx}
\address{Departamento de F\'isica, Universidad Aut\'onoma Metropolitana-Iztapalapa, San Rafael Atlixco 186, Ciudad de M\'exico 09340, M\'exico}
\date{\today}

\begin{abstract}
We develop a theoretical framework based on the cavity and replica methods to analyze the spectral properties of sparse asymmetric correlation matrices of the form $\bm{F} = (\bm{X}\bm{Y}^\top + \omega \bm{Y}\bm{X}^\top)/2T$,  where $\bm{X}$ and $\bm{Y}$ are adjacency matrices of weighted Erd\H{o}s–Rényi random graphs.  We examine how the spectral density evolves as the asymmetry parameter $\omega$ varies from $0 < \omega < 1$ (nearly symmetric matrices) to $-1 < \omega \le 0$ (nearly antisymmetric matrices).  Analytical predictions are validated through exact numerical diagonalization, showing excellent agreement with theoretical results in the thermodynamic limit.
\end{abstract}

\maketitle

\section{Introduction}
The development of random matrix theory arguably began with Wishart’s 1928 work, which investigated the correlation matrices of random variables \cite{Wishart1928}. Since then, a vast range of applications of random matrix theory in statistics and related disciplines has emerged \cite{Mehta2004,Akemann2015}, among which the celebrated Mar\v{c}enko–Pastur problem \cite{Marcenko1967} holds a central place. The results of this study have been extensively used to test whether the variables in a given time series are independent. Specifically, if the variables are independent and identically distributed with zero mean and unit variance, then in the thermodynamic limit the spectral density of their covariance matrix follows the Mar\v{c}enko–Pastur law, regardless of the underlying distribution. Correlations among the variables manifest as deviations from this law.

The Wishart ensemble has thus become a cornerstone of random matrix theory, with applications extending far beyond its original statistical motivation \cite{Forrester_2010,Akemann2015}. In statistics, it underlies methods such as principal component analysis of large datasets \cite{Majumdar_2009,Majumdar_2012}. In physics, Wishart-type matrices appear in diverse contexts, including supersymmetric gauge theories \cite{Demasure_2003}, disordered and interacting systems \cite{PerezCastillo_2018}, entangled quantum states \cite{Majumdar_2008}, quantum chromodynamics \cite{Verbaarschot_2000}, and wireless communications \cite{Telatar_1999}. Beyond these areas, they have also found broad use in finance for risk estimation and portfolio stability \cite{Burda_2005,Laloux_1999}.

A noteworthy generalization of the Mar\v{c}enko–Pastur problem was proposed by Bouchaud and Potters \cite{Bouchaud2020}, who, using techniques from free random matrix theory (see, e.g., \cite{Tulino2004}), studied the spectral density of the symmetric cross-correlation matrix between two time series. They showed that the resulting family of spectral distributions exhibits a rich behavior, interpolating between the Mar\v{c}enko–Pastur law and the Wigner semicircle law. More recently, this framework was extended to diluted random matrices \cite{PerezCastillo2022}, in which the proportion of nonzero entries scales linearly with the matrix size.

In parallel, increasing attention has been devoted to non-Hermitian random matrices, which naturally arise in a variety of settings such as open quantum systems \cite{Rotter_2009,Moiseyev_2011}, biological networks \cite{Allesina_2015}, financial markets \cite{Burda_2004}, and complex dynamical systems \cite{Hatano_1996,Shnerb_1998}. Unlike Hermitian ensembles that describe equilibrium correlations, non-Hermitian ensembles capture directional or asymmetric interactions, dissipation, and stability properties in complex systems. Their spectra are distributed over the complex plane, leading to qualitatively new behaviors absent in the Hermitian case \cite{Ginibre_1965,Akemann2015,Metz_2019}. As a result, non-Hermitian extensions of classical random matrix ensembles are not only mathematically rich but also physically meaningful.

In this work, we extend the results of Refs.~\cite{Bouchaud2020,PerezCastillo2022} by introducing a non-Hermitian parameter $\omega$ (defined precisely below) to study the spectra of diluted and asymmetric correlation matrices. When $\omega = 1$, the model of Ref.~\cite{PerezCastillo2022} is recovered, whereas for $\omega \neq 1$ the theory enables us to explore the complex spectra of asymmetric correlation matrices. This study complements earlier efforts to generalize the Wishart ensemble, encompassing both Hermitian modifications \cite{Akemann_2008} and non-Hermitian extensions \cite{Akemann2011}.

Our approach relies on the cavity and replica methods, originally developed in the study of spin glasses \cite{Mezard1986} and later adapted to random matrix theory. These techniques exploit the connection between random matrices and weighted random graphs, as well as the Edwards–Jones formula \cite{Edwards1976}, to compute the spectral density using tools from statistical physics. Although these approaches have been primarily applied to Hermitian matrices \cite{Nagao_2007,Rogers2008}, with suitable modifications they can also be used to analyze non-Hermitian ensembles \cite{Rogers2009,Neri_2012,Neri_2016,RamosSanchez2020}, making them ideally suited for the present study.

\section{Model definitions}
Consider two $N$-dimensional time series with $T$ time steps, $x_{it}$ and $y_{it}$, where $i = 1, \dots, N$ and $t = 1, \dots, T$. The ratio $q = N/T$ quantifies the relative dimensionality of the data and plays a key role in determining the shape of the eigenvalue distribution of the covariance matrix.

Suppose that the pairs $(x_{it}, y_{it})$ are independent and identically distributed Poisson-like random variables drawn from the distribution
\begin{equation}
P(x_{it}, y_{it}) = \frac{d}{N}\,\rho(x_{it}, y_{it}) + \left(1 - \frac{d}{N}\right)\delta_{x_{it},0}\delta_{y_{it},0},
\end{equation}
where $d/N$ denotes the probability that $(x_{it}, y_{it})$ is nonzero, and $\rho$ is a bivariate distribution with zero mean, unit variance, and correlation coefficient $c = \mathbb{E}(x_{it}y_{it})$. 

Let $\bm{X}$ and $\bm{Y}$ denote the $N \times T$ matrices whose entries are $x_{it}$ and $y_{it}$, respectively. Each of these matrices can be regarded as the weighted adjacency matrix of a bipartite random graph, where a nonzero entry $x_{it}$ (or $y_{it}$) indicates the presence of an edge between node $i = 1, \dots, N$ and node $t = 1, \dots, T$. In this context, $\bm{X}$ and $\bm{Y}$ represent weighted Erd\H{o}s–Rényi random graphs \cite{Erdos1959}, and the parameter $d$ corresponds to their average connectivity. The underlying graph structure of $\bm{X}$ and $\bm{Y}$ is identical; they differ only in the weights assigned to their edges. For this graph, we denote by $\partial i$ the set of nodes $t$ connected to node $i$, and by $\partial t$ the set of nodes connected to node $t$.

In this work, we study the spectral density of asymmetric cross-correlation matrices $\bm{F}$ of the form
\begin{equation}
\bm{F} = \frac{1}{2d}\big(\bm{X}\bm{Y}^\top + \omega \bm{Y}\bm{X}^\top\big),
\label{eq:defF}
\end{equation}
where $\omega \in (-1,1)$ is an asymmetry parameter that interpolates between antisymmetric matrices (for $\omega = -1$) and symmetric Wishart matrices (for $\omega = 1$) \cite{Bishop2018}.

Three representative limits of $\omega$ are of particular interest. For $\omega = 1$, the model coincides with the symmetric correlation matrix studied in Ref.~\cite{PerezCastillo2022}. For $\omega = 0$ and $\omega = -1$, it corresponds to the unsymmetrized and skew-symmetric components of $\bm{X}\bm{Y}^\top$, respectively, as examined in Refs.~\cite{Bouchaud2006,HazarikaPaul2024}, albeit for different choices of $\bm{X}$ and $\bm{Y}$.

\section{Theoretical derivation of the spectral density}
A key property of the Wishart ensemble is that many of its spectral features can be computed analytically \cite{Burda_2005}. Here, we outline how to obtain the spectral density for the generalized asymmetric case using the cavity method. Full details of the derivation appear in Appendix~\ref{app:asymmetric}.

Let $\lambda_{\bm F}^1,\dots,\lambda_{\bm F}^N$ denote the (in general complex) eigenvalues of $\bm F$. The spectral density evaluated at $z=x+iy$ is defined as
\begin{equation}
\rho_{\bm F}(z)=\frac{1}{N}\sum_{i=1}^{N}\delta\!\left(z-\lambda_{\bm F}^i\right).
\end{equation}
Using a slight variation of the Edwards–Jones formula \cite{Edwards1976}, detailed in Appendix~\ref{app:ComplexAnalysis}, this spectral density can be expressed in terms of a \emph{complex partition function} $Z_{\bm F}$ of a system of interacting spinors (i.e.\ elements of $\mathbb{C}^2$), establishing a useful analogy with statistical physics. The final
result is \cite{RamosSanchez2020}
\begin{equation}
\rho_{\bm F}(z)= -\lim_{\eta\to0^+}\frac{1}{N\pi}\,\partial_z^{*}\partial_z \ln Z_{\bm F},
\label{eq:rhoSpinors0}
\end{equation}
where $\partial_z=\tfrac{1}{2}\!\left(\partial_x - i\,\partial_y\right)$ and
$\partial_z^{*}=\tfrac{1}{2}\!\left(\partial_x + i\,\partial_y\right)$ are the Wirtinger derivatives,
and $\eta>0$ is a regularization parameter ensuring convergence of the integral defining $Z_{\bm F}$:
\begin{equation}
Z_{\bm F}=\int d\bm\psi\, e^{-H_{\bm F}}.
\end{equation}
Here $d\bm\psi$ denotes the homogeneous measure over $\mathbb{C}^{2N}$—the configuration space of spinors $\psi_1,\dots,\psi_N$—normalized so that $\int d\bm\psi\,e^{-\bm\psi^\dagger\bm\psi}=1$. The Hamiltonian is
\begin{equation}
H_{\bm F}=
\sum_{i=1}^{N}\psi_i^\dagger \bm M\,\psi_i
\;-\; i\sum_{i,j=1}^{N}\psi_i^\dagger\!\left(F_{ij}\bm\sigma^{+}+F_{ji}^{*}\bm\sigma^{-}\right)\!\psi_j,
\label{eq:HFdef}
\end{equation}
where $F_{ij}$ is the $(i,j)$ entry of $\bm F$, and $\bm M=\eta\,\bm 1_{2}+ i z\,\bm\sigma^{+}+ i z^{*}\,\bm\sigma^{-}$. Here $\bm 1_n$ denotes the $n\times n$ identity matrix and $\bm\sigma^{\pm}$ are the Pauli ladder matrices:
\begin{equation}
\bm\sigma^{+}=\begin{pmatrix}0&1\\[2pt]0&0\end{pmatrix},\qquad
\bm\sigma^{-}=\begin{pmatrix}0&0\\[2pt]1&0\end{pmatrix}.
\end{equation}

Exploiting the analogy with statistical mechanics, differentiating $\ln Z_{\bm F}$ in Eq.~\eqref{eq:rhoSpinors0} gives
\begin{equation}
\rho_{\bm F}(z) =
-\lim_{\eta\to0^+}\frac{i}{N\pi}\,\partial_z^{*}
\sum_{i=1}^{N}\big\langle \psi_i^\dagger \bm\sigma^{+}\psi_i \big\rangle_{H_{\bm F}},
\label{eq:rhoSpinors}
\end{equation}
where $\langle\cdot\rangle_{H_{\bm F}}$ denotes the canonical (thermal) average with respect to $H_{\bm F}$. Since each average involves a single site $i$, it can be written in terms of the single-site marginal $P_i(\psi_i)$:
\begin{equation}
\big\langle \psi_i^\dagger \bm\sigma^{+}\psi_i \big\rangle_{H_{\bm F}}
= \int d\psi_i\, P_i(\psi_i)\,\psi_i^\dagger \bm\sigma^{+}\psi_i,
\label{eq:piDefinition}
\end{equation}
with
\begin{equation}
P_i(\psi)=\frac{1}{Z_{\bm F}}\int d\bm\psi\, e^{-H_{\bm F}}\,\delta(\psi_i-\psi).
\label{eq:piDefinition2}
\end{equation}

For a general matrix $\bm F$, the marginals $P_i$ cannot be computed exactly. However, when the underlying graphs of $\bm X$ and $\bm Y$ are tree-like, they can be obtained via the \emph{cavity method}. The \emph{Bethe approximation} consists in assuming that these relations hold even when the graph is only
approximately tree-like \cite{Peierls1936}. For Erd\H{o}s–Rényi random graphs, this approximation is exact in the thermodynamic limit $N\to\infty$ \cite{Mezard2001}. We briefly outline the computation of the corresponding cavity distributions.

For each node $t$, introduce the spinor pair
\begin{equation}
\mathxy{Y}_{t}=\frac{1}{\sqrt d}\sum_{i\in\partial t} y_{it}\psi_i,
\qquad
\mathxy{X}_{t}=\frac{1}{\sqrt d}\sum_{i\in\partial t} x_{it}\psi_i,
\end{equation}
which allows us to rewrite $H_{\bm F}$ as
\begin{equation}
H_{\bm F}=
\sum_{i=1}^{N}\psi_i^\dagger \bm M\,\psi_i
\;-\; i\sum_{t=1}^{T}\Big(\mathxy{Y}_{t}^\dagger \bm s\, \mathxy{X}_{t}
+ \mathxy{X}_{t}^\dagger \bm s^\dagger \mathxy{Y}_{t}\Big),
\label{eq:HFpsixy}
\end{equation}
where $\bm s=(\omega\,\bm\sigma^{+}+\bm\sigma^{-})/2$.

For each edge $(j,t)$, define the \emph{cavity distribution} $P^{(t)}_{j}(\psi_j)$ as the marginal of $\psi_j$ in the system where node $t$ is removed (equivalently, from the Hamiltonian with all terms containing $t$ deleted). Under the tree-like assumption, once $t$ is removed the neighboring spinors
$\{\psi_i: i\in\partial t\}$ become statistically independent, greatly simplifying the analysis. Similarly, define $P^{(j)}_{t} (\mathxy{X}_{t},\mathxy{Y}_{t})$ as the distribution of $(\mathxy{X}_{t},\mathxy{Y}_{t})$ in the system where node $j$ is removed. Iterating this decoupling gives a closed system for $P^{(t)}_{j}$ and $P^{(j)}_{t}$ for all $j=1,\dots,N$ and $t\in\partial j$.

Assuming Gaussian forms,
\begin{align}
P^{(t)}_{j}(\psi_j) &\propto \exp\!\big[-\psi_j^\dagger \tilde{\bm\Sigma}^{(t)}_{j}\psi_j\big],\\
P^{(j)}_{t}(\mathxy{X}_{t},\mathxy{Y}_{t})
&\propto \exp\!\big[-(\mathxy{X}_{t},\mathxy{Y}_{t})^\dagger
\tilde{\bm\Lambda}^{(t)}_{j}(\mathxy{X}_{t},\mathxy{Y}_{t})\big],
\end{align}
with $\tilde{\bm\Sigma}^{(t)}_{j}\in\mathbb{C}^{2\times2}$ and
$\tilde{\bm\Lambda}^{(t)}_{j}\in\mathbb{C}^{4\times4}$, we obtain the \emph{cavity equations}
(see Appendix~\ref{app:asymmetric}):
\begin{align}
\bm\Lambda^{(i)}_{t}&=
\frac{1}{d}\sum_{j\in\partial t\setminus i}
\begin{pmatrix}
x_{jt}^{2}\tilde{\bm\Sigma}^{(t)}_{j} & y_{jt}x_{jt}\tilde{\bm\Sigma}^{(t)}_{j}\\[0.4em]
y_{jt}x_{jt}\tilde{\bm\Sigma}^{(t)}_{j} & y_{jt}^{2}\tilde{\bm\Sigma}^{(t)}_{j}
\end{pmatrix},
\label{eq:cavEquationsMain0}
\\[4pt]
\big[\tilde{\bm\Sigma}^{(t)}_{j}\big]^{-1}
&= \bm M + \sum_{\tau\in\partial j\setminus t}\bm R_{x_{j\tau}y_{j\tau}}\!\big(\bm W_{\bm\Lambda^{(j)}_{\tau}}\big),
\label{eq:cavEquationsMain}
\end{align}
where $\bm R$ maps $4\times4$ matrices to $2\times2$ matrices via
\begin{equation}
\bm R_{xy}(\bm U)\equiv \frac{1}{d}\Big(x^2\bm U_{[11]} + xy(\bm U_{[21]}+\bm U_{[12]}) + y^2\bm U_{[22]}\Big),
\label{eq:RMainDef}
\end{equation}
with $\bm U_{[ab]}$ the $2\times2$ block $(a,b)$ of $\bm W$, and where, for a $4\times4$ matrix $\bm\Lambda$,
\begin{equation}
\bm W_{\bm\Lambda} =
-i
\begin{pmatrix}
\bm 0_2 & \bm s^\dagger\\
\bm s & \bm 0_2
\end{pmatrix}
\!\left[\bm 1_{4}
- i\,\bm\Lambda
\begin{pmatrix}
\bm 0_2 & \bm s^\dagger\\
\bm s & \bm 0_2
\end{pmatrix}\right]^{-1}\!,
\label{eq:WCavity0}
\end{equation}
with $\bm 0_2$ the $2\times2$ zero matrix.

The single-site marginal in the original system has the form $P_j(\psi_j)\propto \exp[-\psi_j^\dagger \tilde{\bm\Sigma}_{j}\psi_j]$, where
\begin{equation}
\big[\tilde{\bm\Sigma}_{j}\big]^{-1}
= \bm M + \sum_{\tau\in\partial j}\bm R_{x_{j\tau}y_{j\tau}}\!\big(\bm W_{\bm\Lambda^{(j)}_{\tau}}\big).
\label{eq:CiFinal}
\end{equation}
Combining Eqs.~\eqref{eq:rhoSpinors} and \eqref{eq:CiFinal} yields
\begin{equation}
\rho_{\bm F}(z)= -\lim_{\eta\to0^+}\frac{i}{N\pi}\sum_{i=1}^{N}\partial_z^{*}\big[\tilde{\bm\Sigma}_i\big]_{21}.
\label{eq:finalSpectralCavity}
\end{equation}
Taking $\partial_z^{*}$ of Eqs.~\eqref{eq:cavEquationsMain0}–\eqref{eq:cavEquationsMain} leads to a companion set of equations for $\partial_z^{*}\tilde{\bm\Sigma}_i$, reported as Eqs.~(\ref{eq:dStarsC})–(\ref{eq:dStarCavityEquations}) in the appendices.

Although closed forms are generally unavailable, the system can be solved numerically with high precision via fixed-point iterations. Starting from an initial guess for
$\{\bm\Lambda^{(j)}_{t},\,\partial_z^{*}\bm\Lambda^{(j)}_{t},\,\tilde{\bm\Sigma}^{(t)}_{j},\,\partial_z^{*}\tilde{\bm\Sigma}^{(t)}_{j}\}$, iterating Eqs.~(\ref{eq:cavEquationsMain0})–(\ref{eq:cavEquationsMain}) and (\ref{eq:dStarsC})–(\ref{eq:dStarCavityEquations}) converges to accurate fixed points, allowing one to compute the spectral density of a typical matrix $\bm F$ in the ensemble.

Alternatively, ensemble averaging the cavity equations yields the average spectral density, which is equivalent to the replica approach (see Appendix~\ref{app:replicaCavity} for details). The result is
\begin{equation}
\big\langle \rho(z) \big\rangle_{\bm F}=
\frac{i}{\pi}\int d\tilde{\bm\Sigma}\, d\tilde{\bm\Sigma}^{\star}\,
\tilde{\omega}(\tilde{\bm\Sigma},\tilde{\bm\Sigma}^{\star})\,
\tilde{\Sigma}^{\star}_{21},
\label{eq:finalRhoDeltaReplica}
\end{equation}
where $\tilde{\omega}(\tilde{\bm\Sigma},\tilde{\bm\Sigma}^{\star})$ is a probability density over pairs of $2\times2$ complex matrices. It satisfies the following self-consistent system involving two auxiliary distributions $\hat{\omega}(\hat{\bm\Gamma},\hat{\bm\Gamma}^{\star})$ and $m(\bm\Lambda,\bm\Lambda^{\star})$, with $\hat{\bm\Gamma},\hat{\bm\Gamma}^{\star}\in\mathbb{C}^{2\times2}$ and
$\bm\Lambda,\bm\Lambda^{\star}\in\mathbb{C}^{4\times4}$:
\begin{equation}
\begin{split}
\hat{\omega}(\hat{\bm\Gamma},\hat{\bm\Gamma}^{\star})
&=\int d\bm\Lambda\, m(\bm\Lambda,\bm\Lambda^{\star})
\bigg\langle
\delta\!\Big(\hat{\bm\Gamma}-\bm R_{xy}[\bm W_{\bm\Lambda}]\Big)\,
\delta\!\Big(\hat{\bm\Gamma}^{\star}+\bm R_{xy}[\bm W_{\bm\Lambda}\bm\Lambda^{\star}\bm W_{\bm\Lambda}]\Big)
\bigg\rangle_{\rho},
\\[4pt]
\tilde{\omega}(\tilde{\bm\Sigma},\tilde{\bm\Sigma}^{\star})
&=\sum_{k=0}^{\infty}\frac{e^{-d/q}d^{k}}{q^{k}k!}
\int\!\prod_{\lambda=1}^{k} d\hat{\bm\Gamma}_{\lambda}\, d\hat{\bm\Gamma}^{\star}_{\lambda}\,
\hat{\omega}(\hat{\bm\Gamma}_{\lambda},\hat{\bm\Gamma}^{\star}_{\lambda})
\\[-2pt]&\quad\times
\delta\!\Big(\tilde{\bm\Sigma}-\big[\sum_{\lambda=1}^{k}\hat{\bm\Gamma}_{\lambda}+\bm M\big]^{-1}\Big)\,
\delta\!\Big(\tilde{\bm\Sigma}^{\star}+\tilde{\bm\Sigma}
\big[\sum_{\lambda=1}^{k}\hat{\bm\Gamma}^{\star}_{\lambda}+ i\,\bm\sigma_{-}\big]\tilde{\bm\Sigma}\Big),
\\[4pt]
m(\bm\Lambda,\bm\Lambda^{\star})
&=\sum_{k=0}^{\infty} p_k
\int\!\prod_{\lambda=1}^{k} d\tilde{\bm\Sigma}_{\lambda}\,
\tilde{\omega}(\tilde{\bm\Sigma}_{\lambda},\tilde{\bm\Sigma}^{\star}_{\lambda})
\\[-2pt]&\quad\times
\Bigg\langle
\delta\!\Big(\bm\Lambda-\frac{1}{d}\sum_{\lambda=1}^{k}
\begin{pmatrix}
x_{\lambda}^{2}\tilde{\bm\Sigma}_{\lambda} & x_{\lambda}y_{\lambda}\tilde{\bm\Sigma}_{\lambda}\\
x_{\lambda}y_{\lambda}\tilde{\bm\Sigma}_{\lambda} & y_{\lambda}^{2}\tilde{\bm\Sigma}_{\lambda}
\end{pmatrix}\Big)
\\[-2pt]&\qquad\qquad\times
\delta\!\Big(\bm\Lambda^{\star}-\frac{1}{d}\sum_{\lambda=1}^{k}
\begin{pmatrix}
x_{\lambda}^{2}\tilde{\bm\Sigma}^{\star}_{\lambda} & x_{\lambda}y_{\lambda}\tilde{\bm\Sigma}^{\star}_{\lambda}\\
x_{\lambda}y_{\lambda}\tilde{\bm\Sigma}^{\star}_{\lambda} & y_{\lambda}^{2}\tilde{\bm\Sigma}^{\star}_{\lambda}
\end{pmatrix}\Big)
\Bigg\rangle_{\rho^{k}},
\end{split}
\label{eq:saddleCompleteFamily}
\end{equation}
where $q=N/T$, $\bm R_{xy}$ and $\bm W_{\bm\Lambda}$ are given by Eqs.~\eqref{eq:RMainDef}–\eqref{eq:WCavity0}, $\langle\cdot\rangle_{\rho}$ denotes averaging with respect to $\rho(x,y)$, and $\langle\cdot\rangle_{\rho^{k}}$ averages over $k$ independent draws $(x_{\lambda},y_{\lambda})\sim\rho$.

These equations can be solved using population dynamics \cite{Mezard2001}; implementation details can be found in \cite{Kuhn_2008,Metz_2016}.

\section{Comparison with numerical results}
We assess the validity of our theoretical predictions by comparing the average spectral density obtained from Eq.~\eqref{eq:finalRhoDeltaReplica} with results from direct numerical diagonalization. To illustrate the versatility of the model and its ability to capture distinct behaviors, we consider three representative regimes. In the \emph{nearly anti-Hermitian (skew-Hermitian) regime}, where the spectrum concentrates along the imaginary axis, we set $\omega=-0.8$, $q=1$, and $c=0.8$. In the \emph{nearly Hermitian regime}, where eigenvalues cluster near the real axis, we use $\omega=0.3$, $q=1$, and $c=0.3$. Finally, in the \emph{mixed regime}, where eigenvalues are distributed along both axes with comparable weight, we take $\omega=-0.2$, $q=1/2$, and $c=0.1$. In all cases, the distribution $\rho(x,y)$ is a bivariate Gaussian with correlation coefficient $c$, and the underlying graphs have average connectivity $d=5$.

For the theoretical curves, Eq.~\eqref{eq:saddleCompleteFamily} is solved numerically using the population dynamics algorithm with population size $10^4$, regularizer $\eta=10^{-8}$, and $5\times10^4$ updates. For the direct diagonalizations, we sample $300$ matrices of size $N=20\,000$, yielding $6\times10^6$ eigenvalues per regime.

\begin{figure}[h]
\centering
\includegraphics[scale=0.25]{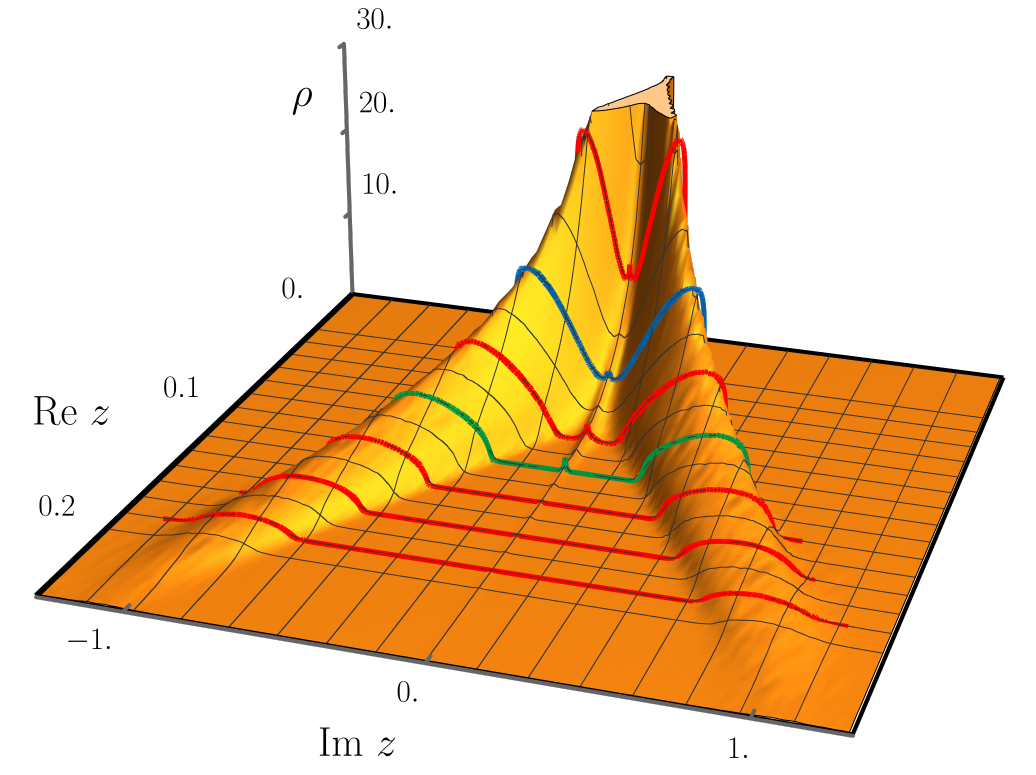}
\includegraphics[scale=0.35]{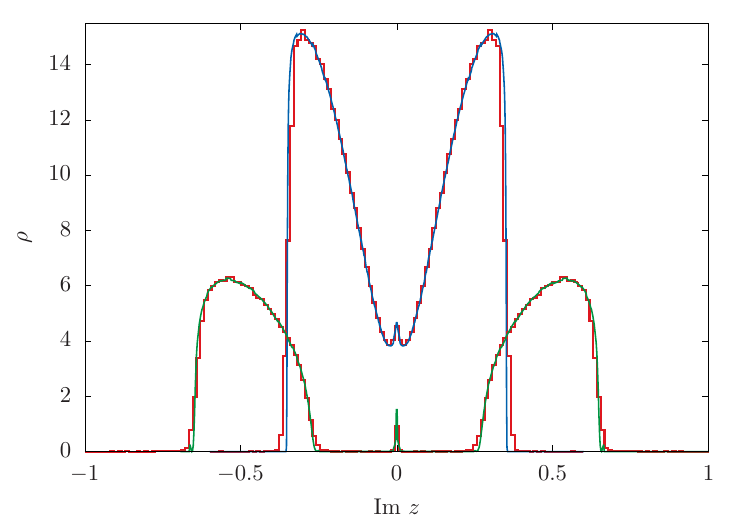}
\includegraphics[scale=0.25]{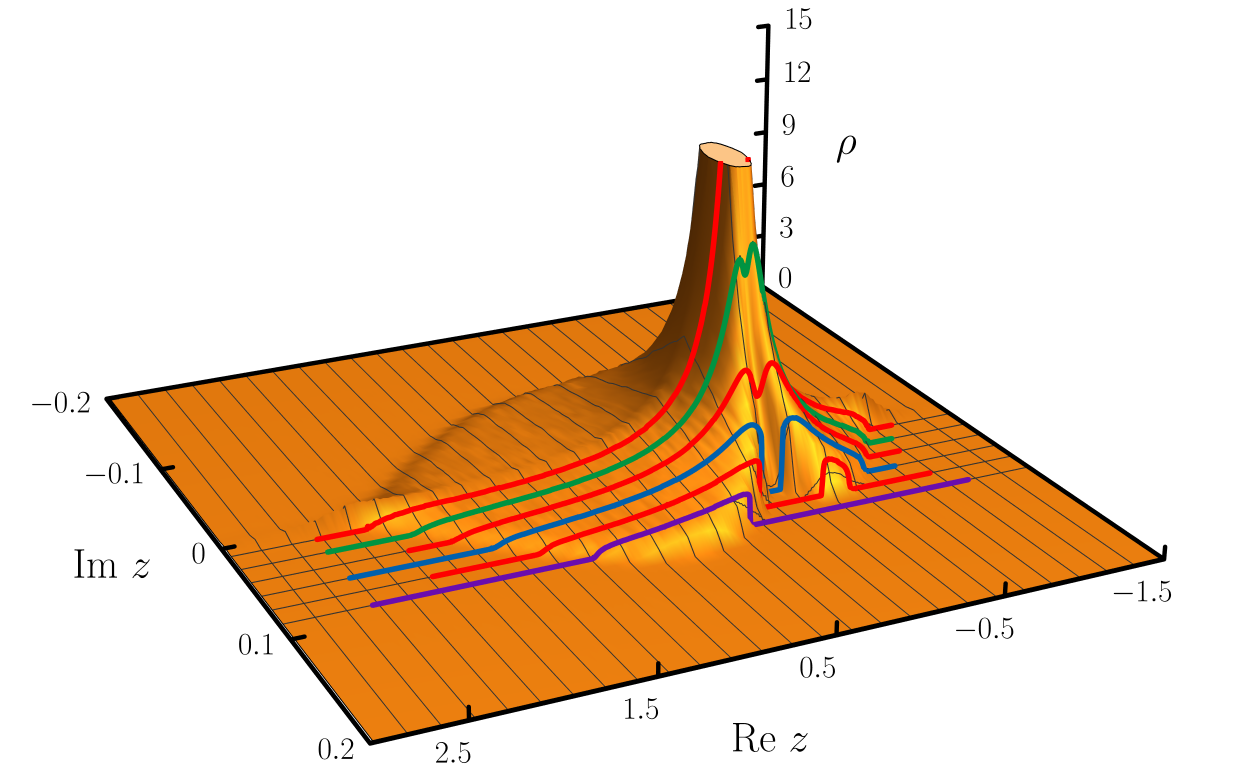}
\includegraphics[scale=0.35]{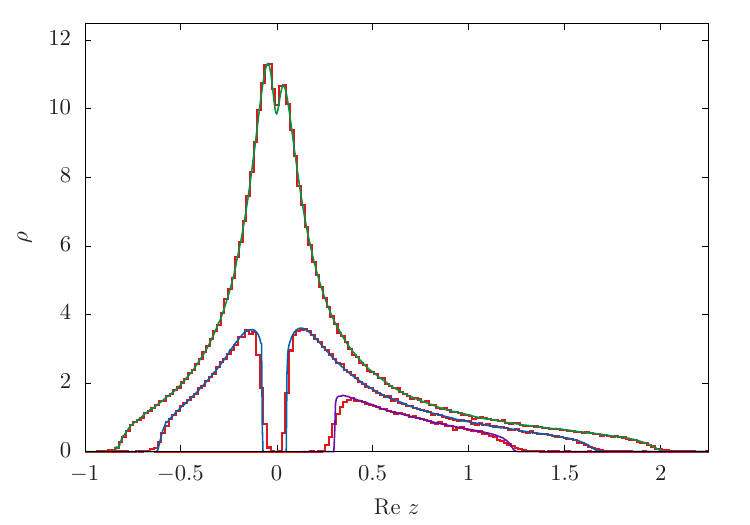}
\includegraphics[scale=0.25]{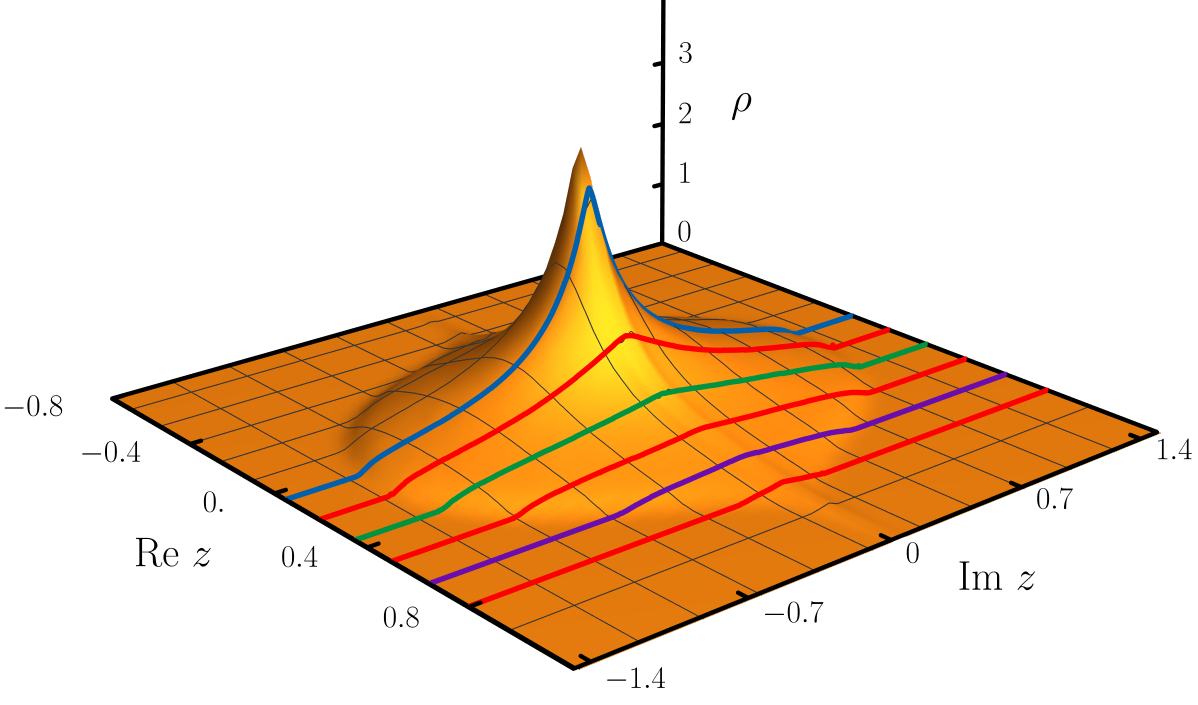}
\includegraphics[scale=0.35]{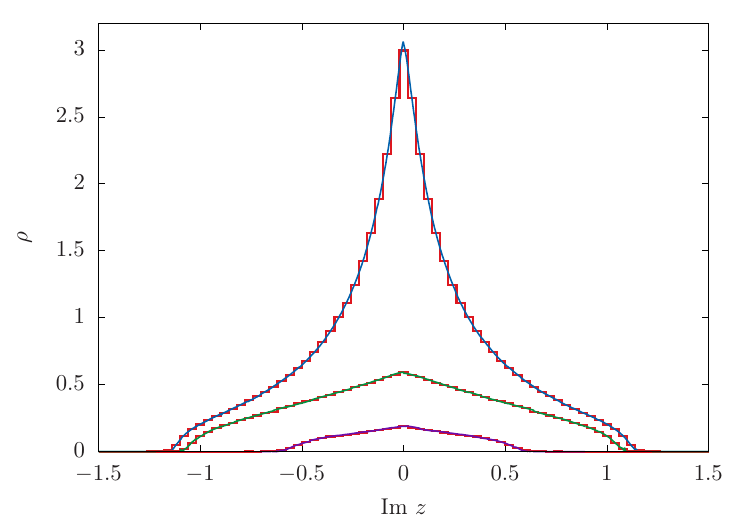}
\caption{Average spectral density of $\bm F$ from Eq.~\eqref{eq:defF} for different parameter choices. First row: $q=1$, $d=5$, $c=0.8$, $\omega=-0.8$; second row: $q=1$, $d=5$, $c=0.3$, $\omega=0.3$; third row: $q=1/2$, $d=5$, $c=0.1$, $\omega=-0.2$. Left panels: surfaces from numerical diagonalization (points) with theoretical predictions overlaid as solid coloured curves. Right panels: cuts of the spectral density along specific directions—first row: $\re\,z=0.075$ (blue), $\re\,z=0.125$ (green); second row: $\im\,z=0.025$ (green), $\im\,z=0.055$ (blue), $\im\,z=0.085$ (purple); third row: $\re\,z=0.05$ (blue), $\re\,z=0.35$ (green), $\re\,z=0.65$ (purple). In all cases, $300$ matrices are sampled, yielding $6\times10^6$ eigenvalues. Matrix sizes: $N=T=20\,000$ for the first and second rows; $N=T/2=20\,000$ for the third row.}
\label{fig:omegaComparisons}
\end{figure}

\begin{figure}[h]
\centering
\includegraphics[scale=0.35]{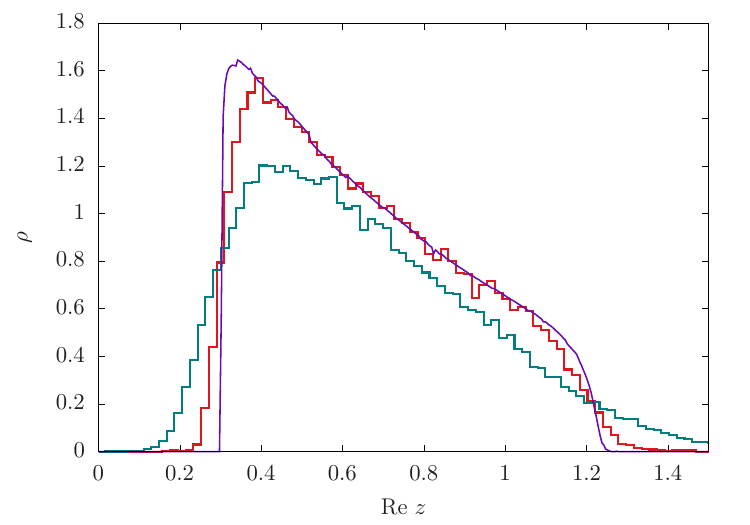}
\caption{Finite-size effects in the spectral density for the parameters of the second row of Fig.~\ref{fig:omegaComparisons}, along the cut $\im\,z=0.085$. Red: numerical diagonalization with $40\,000\times40\,000$ matrices; green: $1000\times1000$ matrices; purple: theoretical prediction. The agreement improves with increasing $N$, consistent with convergence to the thermodynamic limit.}
\label{fig:finiteSizeComparison}
\end{figure}

As shown in Fig.~\ref{fig:omegaComparisons}, the agreement between theory and numerics is excellent. The small discrepancies near the spectral edges are attributable to finite-size effects. This is confirmed in Fig.~\ref{fig:finiteSizeComparison}, which shows that these differences diminish as $N$ increases. A key practical advantage of our approach is that it provides direct access to the thermodynamic-limit spectral density, avoiding the computational cost of diagonalizing extremely large matrices.

The results also reveal several structural features. Even for a relatively modest value such as $\omega=0.3$, the eigenvalues are almost entirely real, producing a spectral density close to that reported in Ref.~\cite{PerezCastillo2022}. Moreover, among the three cases considered, the mixed regime with $q=1/2$ is the only one in which the spectral density remains bounded at the origin. This aligns with Ref.~\cite{PerezCastillo2022}, where decreasing $q$ was shown to change the spectrum from unbounded to bounded, and further illustrates how $\omega$, $q$, and $c$ jointly shape the spectral properties.

\section{Conclusions} 
We have analyzed the spectral density of a family of sparse, asymmetric cross–correlation matrices of the form in Eq.~\eqref{eq:defF} using the cavity and replica methods. In the limit $\omega=1$ we recover the Hermitian (symmetric) case studied previously, while for $\omega\neq 1$ we uncover a range of genuinely non-Hermitian regimes whose spectral features can be treated systematically within our framework. We validated the theoretical predictions by large-scale numerical diagonalization across representative parameter choices, observing excellent agreement and clarifying finite-size effects near the spectral edges.

More broadly, the approach provides a flexible route to non-Hermitian generalizations of the Wishart ensemble. It directly accesses the thermodynamic-limit density without resorting to diagonalization, and it can be extended to incorporate additional structural ingredients (e.g., different dilution levels, degree heterogeneity, or alternative edge–weight distributions). Understanding how the spectral density encodes the statistical structure and cross–correlations of time series thus becomes amenable to a controlled analysis in sparse, asymmetric settings.

\setcounter{equation}{0}
\setcounter{section}{0}
\setcounter{figure}{0}
\setcounter{table}{0}
\makeatletter
\renewcommand{\theequation}{S\arabic{equation}}
\renewcommand{\thefigure}{S\arabic{figure}}
\renewcommand{\thesection}{S\arabic{section}}

\FloatBarrier
\begin{center}
{\large\textbf{Appendices}}\\
\end{center}
In these appendices we present:
\begin{enumerate}
\item A simple derivation of the Edwards–Jones formula using differential forms in complex analysis.
\item Details of the cavity–method computation of the spectral density for a typical matrix $\bm F$ in the ensemble.
\item The computation of the ensemble–average spectral density via the replica method.
\item A proof of the equivalence between the two approaches.
\end{enumerate}

\section{A simple derivation of the Edwards–Jones formula}
\label{app:ComplexAnalysis}
We begin with some preliminaries from complex analysis. For a complex function $f(z,z^*)$, its exterior derivative is
\begin{equation}
df \;=\; \partial_z f \, dz \;+\; \partial_{z}^* f \, d z^*.
\end{equation}
Let $\Omega$ be a region in the complex plane. By Stokes’ theorem,
\begin{equation}
\oint_{\partial \Omega} \frac{1}{z} \, dz
= \int_\Omega d \!\left(\frac{dz}{z}\right)
= \int_\Omega \partial_{z} \!\left(\frac{1}{z}\right) dz\wedge dz+
 \int_\Omega \partial^*_{z} \!\left(\frac{1}{z}\right) dz^* \wedge dz
 =\int_\Omega \partial^*_{z} \!\left(\frac{1}{z}\right) dz^* \wedge dz,
\end{equation}
where the first term vanishes because $dz \wedge dz = 0$.

Using this identity together with the \emph{Residue} theorem, we obtain
\begin{equation}
\oint_{\partial\Omega} \frac{1}{z} \, dz
= \int_\Omega \partial^*_{z} \!\left(\frac{1}{z}\right) dz^* \wedge dz
= \begin{cases}
2\pi i & 0 \in \Omega, \\[4pt]
0 & 0 \notin \Omega,
\end{cases}
\end{equation}
which implies
\begin{equation}
\partial^*_{z}\!\left(\frac{1}{z}\right) \;=\; \pi\,\delta(x)\,\delta(y),
\label{eq:dzdirac}
\end{equation}
with $z=x+iy$, $z^*=x-iy$, and $dz^* \wedge dz = 2i\,dx\wedge dy$.

From \eqref{eq:dzdirac} it follows immediately that
\begin{equation}
\partial^*_{z}\,\partial_{z} \ln z \;=\; \pi\,\delta(x)\,\delta(y).
\end{equation}
It will be convenient to reverse the order of differentiation. Note that
\begin{equation}
\partial^*_{z} \ln z \;=\; -\partial^*_{z} \ln\!\left(\frac{1}{z}\right)
= -\,z\,\partial^*_{z}\!\left(\frac{1}{z}\right)
= -\,\pi\,z\,\delta(x)\,\delta(y) \;=\; 0,
\end{equation}
so that $\partial_{z}\,\partial^*_{z} \ln z = 0$.

Writing $\ln z = \ln|z| + i\theta$, where $\theta$ is the phase of $z$, we have
\begin{equation}
\begin{split}
\partial^*_{z}\,\partial_{z}\big(\ln|z| + i\theta\big) &= \pi\,\delta(x)\,\delta(y), \\
\partial_{z}\,\partial^*_{z}\big(\ln|z| + i\theta\big) &= 0.
\end{split}
\end{equation}
Adding and subtracting these relations, and using that
$\partial^*_{z}\,\partial_{z}+\partial_{z}\,\partial^*_{z}$ and
$i(\partial^*_{z}\,\partial_{z}-\partial_{z}\,\partial^*_{z})$ are real operators, yields
\begin{equation}
(\partial^*_{z}\,\partial_{z}+\partial_{z}\,\partial^*_{z}) \ln|z| \;=\; \pi\,\delta(x)\,\delta(y),
\qquad
(\partial^*_{z}\,\partial_{z}-\partial_{z}\,\partial^*_{z}) \ln|z| \;=\; 0.
\end{equation}
Hence
\begin{equation}
\partial^*_{z}\,\partial_{z}\ln(|z|^{2})
\;=\; 2\,\partial^*_{z}\,\partial_{z}\ln|z|
\;=\; \pi\,\delta(x)\,\delta(y).
\label{eq:diracDeltalog}
\end{equation}

Let $\lambda^i_{\bm F}$, $i=1,\dots,N$, be the eigenvalues of an $N\times N$ non-Hermitian matrix $\bm F$.
Define the potential
\begin{equation}
\varphi_{\bm F}(z,z^*) \;=\; \frac{1}{N}\sum_{i=1}^N \log\!\big[(z-\lambda_{\bm F}^i)(z-\lambda_{\bm F}^i)^*\big].
\end{equation}
Using \eqref{eq:diracDeltalog} we find
\begin{equation}
\frac{1}{\pi}\,\partial^*_z\partial_z\,\varphi_{\bm F}(z,z^*)
\;=\; \frac{1}{\pi}\,\frac{1}{N}\sum_{i=1}^N
\partial^*_z\partial_z \log\!\big[(z-\lambda_{\bm F}^i)(z-\lambda_{\bm F}^i)^*\big]
\;=\; \rho_{\bm F}(z,z^*),
\end{equation}
where
\begin{equation}
\rho_{\bm F}(z,z^*) \;=\; \frac{1}{N}\sum_{i=1}^N
\delta\!\big(x-\Re[\lambda_{\bm F}^i]\big)\,\delta\!\big(y-\Im[\lambda_{\bm F}^i]\big)
\end{equation}
is the spectral density.

To express $\varphi_{\bm F}(z,z^*)$ in terms of $\bm F$, note that
\begin{equation}
\varphi_{\bm F}(z,z^*) \;=\; \frac{1}{N}\ln\det\!\Big[(z\bm 1_N-\bm F)\,(z\bm 1_N-\bm F)^\dagger\Big].
\end{equation}
This determinant can be rewritten as
\begin{equation}
\det\!\Big[(z\bm 1_N-\bm F)\,(z\bm 1_N-\bm F)^\dagger\Big]
= \lim_{\eta\to0^+}
\det
\begin{pmatrix}
\eta \bm 1_N & i (z\bm 1_N-\bm F) \\
i (z\bm 1_N-\bm F)^\dagger & \eta \bm 1_N
\end{pmatrix}.
\end{equation}
The regularizer $\eta>0$ is introduced so that the determinant admits a convergent Gaussian
representation:
\begin{equation}
\det
\begin{pmatrix}
\eta \bm 1_N & i (z\bm 1_N-\bm F) \\
i (z\bm 1_N-\bm F)^\dagger & \eta \bm 1_N
\end{pmatrix}
=
\Bigg(\int d\bm\psi\, e^{-H_{\bm F}}\Bigg)^{-1},
\end{equation}
where the measure $d\bm\psi=\prod_{i=1}^N du_i\,du_i^*\,dv_i\,dv_i^*$ runs over $N$ spinors
$\psi_i=(u_i,v_i)$ with $u_i,v_i\in\mathbb{C}$, normalized by
\begin{equation}
\int d\bm\psi\, e^{-\sum_{i=1}^N \psi_i^\dagger \psi_i}=1,
\end{equation}
and $H_{\bm F}$ is the quadratic form
\begin{equation}
H_{\bm F}=
\sum_{i=1}^{N} \psi_{i}^{\dagger} \bm M\,\psi_{i}
\;-\; i\sum_{i,j=1}^{N} \psi_{i}^{\dagger}\big(F_{ij}\bm{\sigma}^{+}+F_{ji}^{*}\bm{\sigma}^{-}\big)\psi_{j},
\end{equation}
with
\begin{equation}
\bm M=
\begin{pmatrix}
\eta & i z \\
i z^* & \eta
\end{pmatrix},\qquad
\bm \sigma^{+}=
\begin{pmatrix}
0 & 1 \\
0 & 0
\end{pmatrix},\qquad
\bm \sigma^{-}=
\begin{pmatrix}
0 & 0 \\
1 & 0
\end{pmatrix}.
\end{equation}
Hence,
\begin{equation}
\rho_{\bm F}(z,z^*)
\;=\; -\lim_{\eta\to0^+}\frac{1}{N\pi}\,\partial^*_z \partial_z
\ln \int d\bm\psi\, e^{- H_{\bm F}}.
\end{equation}

\section{Computation of the spectral density using the cavity method}
\label{app:asymmetric}
In this appendix we present the details needed to derive the cavity equations for the matrix
$\bm F$ of the non-Hermitian Wishart ensemble.

Our starting point is the Hamiltonian $H_{\bm F}$ written as in Eq.~\eqref{eq:HFpsixy}. Fix a node $i$ and
split $H_{\bm F}$ into three parts: the local contribution at $i$, the terms that couple $i$ to its
neighbors $t\in\partial i$, and the remainder that does not involve $i$,
\begin{equation*}
H_{\bm F} \;=\; H_{i} \;+\; \sum_{t \in \partial i} H_{i}^{t}\!\big(\psi_i,\mathxy{X}^{(i)}_{t}[\psi], \mathxy{Y}^{(i)}_{t}[\psi]\big)
\;+\; H^{(i)},
\end{equation*}
where
\begin{align}
H_{i}&=\psi_{i}^{\dagger}\big(\eta \bm{1}_{2}+ i z \bm \sigma^{+}+i z^{*} \bm \sigma^{-}\big)\psi_{i},
\nonumber
\\
H_{i}^{t}\!\big(\psi_i,\mathxy{X}^{(i)}_{t}[\psi], \mathxy{Y}^{(i)}_{t}[\psi]\big)
&=-\Big(\mathxy{Y}_{t}^{(i)}[\psi]+\tfrac{y_{it}}{\sqrt d}\,\psi_{i}\Big)^{\!\dagger}\bm s
\Big(\mathxy{X}^{(i)}_{t}[\psi]+\tfrac{x_{it}}{\sqrt d}\,\psi_{i}\Big)
\;+\;\mathrm{h.c.},
\label{eq:HiDef}
\\
H^{(i)}&=H_{\bm F}- H_{i} - \sum_{t \in \partial i} H_{i}^{t}\!\big(\psi_i,\mathxy{X}^{(i)}_{t}[\psi], \mathxy{Y}^{(i)}_{t}[\psi]\big),
\nonumber
\end{align}
with
\begin{equation}
\mathxy{Y}_{t}^{(i)}[\psi]=\frac{1}{\sqrt d} \sum_{j \in \partial t\setminus i} y_{jt}\psi_{j},
\qquad
\mathxy{X}_{t}^{(i)}[\psi]=\frac{1}{\sqrt d} \sum_{j \in \partial t\setminus i} x_{jt}\psi_{j}.
\end{equation}
Thus $H^{(i)}$ is obtained from $H_{\bm F}$ by removing all interactions that involve node $i$; we
use the superscript $(i)$ to indicate observables in this modified system.

Using these definitions, the marginal $P_i$ from Eq.~\eqref{eq:piDefinition2} can be written as
\begin{equation}
\begin{split}
P_{i}(\psi_{i})&= \frac{1}{Z_{\bm F}}\, e^{-H_i}
\int  \Biggl[\prod_{t \in \partial i} d \mathxy{X}_{t}^{(i)}\, d \mathxy{Y}_{t}^{(i)} \Biggr]
\exp\!\Bigl[-\textstyle\sum_{t \in \partial i} H_{i}^{t}\!\big(\psi_i,\mathxy{X}_{t}^{(i)},\mathxy{Y}_{t}^{(i)}\big)\Bigr]
\\[-2pt]&\quad\times
\int \Biggl[\prod_{t \in \partial i}\prod_{j \in \partial t\setminus i} d \psi_{j} \Biggr]
\prod_{t \in \partial i} \bigl[\delta\big( \mathxy{X}^{(i)}_{t}-\mathxy{X}^{(i)}_{t}[\psi]\big)\,
\delta\big( \mathxy{Y}^{(i)}_{t}-\mathxy{Y}^{(i)}_{t}[\psi]\big)\bigr]
\int \Biggl[\sideset{}{'}\prod_{k} d \psi_{k} \Biggr] e^{-H^{(i)}},
\end{split}
\end{equation}
where the Dirac deltas enforce the definitions of $\mathxy{X}^{(i)}_{t}$ and $\mathxy{Y}^{(i)}_{t}$ in
terms of the neighboring spinors, and the primed product runs over those $\psi_k$ not
appearing in the previous integral (i.e., such that $\partial i \cap \partial k = \varnothing$).

Under the Bethe (locally tree-like) approximation, any correlation among the nodes
$t\in\partial i$ arises only through their common interaction with $i$. Since these
interactions are contained in $\sum_{t\in\partial i} H_i^t$, the second-line integral factorizes over
$t$, yielding
\begin{equation}
\label{eq:PiFromCavities}
P_{i}(\psi_{i})= \frac{e^{-H_i}}{Z^{i}}\,
\prod_{t \in \partial i}\Biggl[
\int  d \mathxy{X}_{t}^{(i)}\, d \mathxy{Y}_{t}^{(i)}\;
e^{- H_{i}^{t}\!\big(\psi_i,\mathxy{X}_{t}^{(i)},\mathxy{Y}_{t}^{(i)}\big)}
\; P_{t}^{(i)}\!\big(\mathxy{X}_{t}^{(i)}, \mathxy{Y}_{t}^{(i)}\big)
\Biggr],
\end{equation}
where $Z^{i}$ is a normalization factor and $P_{t}^{(i)}$ is the \emph{cavity distribution}, i.e.,
the marginal of $\mathxy{X}_{t}^{(i)},\mathxy{Y}_{t}^{(i)}$ in the system where node $i$ is removed:
\begin{equation*}
P_{t}^{(i)}\!\big(\mathxy{X}_{t}^{(i)},\mathxy{Y}_{t}^{(i)}\big)
=\frac{1}{Z_t^{(i)}}
\int \Biggl[\prod_{j \in \partial t\setminus i} d \psi_{j}\,
\sideset{}{'}\prod_{k} d \psi_{k}\Biggr]
 \delta\big( \mathxy{X}^{(i)}_{t}-\mathxy{X}^{(i)}_{t}[\psi]\big)\,
 \delta\big( \mathxy{Y}^{(i)}_{t}-\mathxy{Y}^{(i)}_{t}[\psi]\big)\,
 e^{-H^{(i)}}.
\end{equation*}

Applying the same reasoning to $P_{t}^{(i)}$ (noting that $(\mathxy{X}^{(i)}_{t},\mathxy{Y}^{(i)}_{t})$
does not couple directly to any $\psi_j \neq \psi_i$; cf. Eq.~\eqref{eq:HiDef}), we obtain
\begin{equation}
\label{eq:orsystem1}
P_{t}^{(i)}\!\big(\mathxy{X}_{t}^{(i)},\mathxy{Y}_{t}^{(i)}\big)
=
\frac{1}{Z^{(i,t)}}
\int \Biggl(\prod_{j\in \partial t \setminus i} d\psi_{j}\; P^{(i,t)}_{j}(\psi_{j})\Biggr)
 \delta\big( \mathxy{X}^{(i)}_{t}-\mathxy{X}^{(i)}_{t}[\psi]\big)\,
 \delta\big( \mathxy{Y}^{(i)}_{t}-\mathxy{Y}^{(i)}_{t}[\psi]\big),
\end{equation}
where $P^{(i,t)}_{j}$ is the marginal in the system with nodes $i$ and $t$ removed.
On a tree, removing $i$ and then $t$ is equivalent (for $j\in\partial t\setminus i$) to removing $t$
only, so we write $P^{(i,t)}_{j}\equiv P^{(t)}_{j}$. Repeating the factorization step for $P^{(t)}_{j}$
yields
\begin{equation}
\label{eq:orsystem2}
P^{(t)}_{j}(\psi_{j})= \frac{e^{-H_j}}{Z_j^{(t)}}
\prod_{\tau \in \partial j\setminus t } \Biggl[\int  d \mathxy{X}_{\tau}^{(j)}\, d \mathxy{Y}_{\tau}^{(j)}\;
e^{- H_{j}^{\tau}\!\big(\mathxy{X}_{\tau}^{(j)},\mathxy{Y}_{\tau}^{(j)}\big)}\;
P_{\tau}^{(j)}\!\big(\mathxy{X}_{\tau}^{(j)},\mathxy{Y}_{\tau}^{(j)}\big)\Biggr],
\end{equation}
where $P_{\tau}^{(j)}$ is the marginal in the system with node $j$ removed.
Equations~\eqref{eq:orsystem1}–\eqref{eq:orsystem2} form a closed system for $P_{t}^{(i)}$ and
$P_{j}^{(t)}$.

Assuming Gaussian forms,
\begin{equation}
\label{eq:gaussianSol}
P^{(t)}_{j}(\psi_{j})= \frac{\exp\!\big(- \psi_{j}^{\dagger} [\tilde{\bm{\Sigma}}^{(t)}_{j}]^{-1} \psi_{j}\big)}
{\det \tilde{\bm{\Sigma}}^{(t)}_{j}},
\qquad
P_{\tau}^{(j)}\big(\mathxy{Z}^{(j)}_\tau\big)=
\frac{\exp\!\Big(- \big(\mathxy{Z}^{(j)}_{\tau}\big)^{\dagger}
[\bm \Lambda^{(j)}_{\tau}]^{-1}\mathxy{Z}^{(j)}_{\tau}\Big)}
{\det \bm \Lambda^{(j)}_{\tau}},
\end{equation}
with $\mathxy{Z}^{(j)}_\tau=(\mathxy{X}^{(j)}_{\tau},\mathxy{Y}^{(j)}_{\tau})$,
$\tilde{\bm{\Sigma}}^{(t)}_{j}\in\mathbb{C}^{2\times2}$ and
$\bm \Lambda^{(j)}_{\tau}\in\mathbb{C}^{4\times4}$,
direct substitution into \eqref{eq:orsystem1}–\eqref{eq:orsystem2} and Gaussian integration yield
\begin{equation}
\label{eq:deltaCavity}
\bm \Lambda^{(i)}_{t}=
\frac{1}{d}\sum_{j \in \partial t \setminus i}
\begin{pmatrix}
 x_{jt}^{2}\tilde{\bm{\Sigma}}^{(t)}_{j} & y_{jt} x_{jt}\tilde{\bm{\Sigma}}^{(t)}_{j} \\[0.4em]
 y_{jt} x_{jt}\tilde{\bm{\Sigma}}^{(t)}_{j} & y_{jt}^{2}\tilde{\bm{\Sigma}}^{(t)}_{j}
\end{pmatrix},
\end{equation}
and
\begin{equation}
\label{eq:cCavity}
\big[\tilde{\bm{\Sigma}}^{(t)}_{j}\big]^{-1}
= \bm M+
 \sum_{\tau \in \partial j \setminus t } \bm R_{x_{j\tau}y_{j\tau}}\!\big(\bm W_{\bm \Lambda_\tau^{(j)}}\big),
\end{equation}
where
\begin{equation}
\label{eq:WCavity}
\bm W_{\bm \Lambda^{(j)}_\tau}
=
i \bm S\Big[\bm{1}_{4}+ i\,\bm \Lambda^{(j)}_{\tau}\bm S\Big]^{-1},
\qquad
\bm S=
-\begin{pmatrix}
 \bm 0_2 & \bm s^{\dagger} \\
 \bm s & \bm 0_2
\end{pmatrix},
\end{equation}
$\bm 0_2$ is the $2\times 2$ zero matrix, and $\bm{R}$ maps $4\times4$ to $2\times2$ blocks via
\[
\bm{R}_{xy}(\bm{U})\equiv \frac{1}{d}\Big(x^2 \bm{U}_{[11]}+ xy (\bm{U}_{[21]}+\bm{U}_{[12]}) + y^2 \bm{U}_{[22]}\Big),
\]
with $\bm U_{[ab]}$ the $2\times2$ block $(a,b)$ of $\bm U$.

This system can be solved efficiently by fixed-point iteration: start from initial guesses for
$\bm \Lambda^{(j)}_{\tau}$ and $\tilde{\bm{\Sigma}}^{(t)}_{j}$ and iterate until a desired tolerance is
reached.

Within this framework, the single-site marginal $P_{j}(\psi_{j})$ follows from
\eqref{eq:PiFromCavities} and \eqref{eq:gaussianSol} as
\begin{equation}
P_{j}(\psi_{j})= \frac{\exp\!\big(- \psi_{j}^{\dagger} [\tilde{\bm{\Sigma}}_{j}]^{-1} \psi_{j}\big)}
{\det \tilde{\bm{\Sigma}}_{j}},
\end{equation}
where
\begin{equation}
\label{eq:tildeSigmainv}
\big[\tilde{\bm{\Sigma}}_{j}\big]^{-1}
= \bm M+
\sum_{\tau \in \partial j}\bm R_{x_{j\tau}y_{j\tau}}\!\big(\bm W_{\bm \Lambda_\tau^{(j)}}\big).
\end{equation}
Hence the spectral density \eqref{eq:rhoSpinors}–\eqref{eq:piDefinition} reads
\begin{equation}
\label{eq:rhozCi}
\rho_{\bm F}(z)= -\lim_{\eta \to 0^+}\frac{i}{N \pi}
 \sum_{j=1}^{N} \partial_z^{*}\big[\tilde{\bm{\Sigma}}_{j}\big]_{21}.
\end{equation}

Differentiating \eqref{eq:tildeSigmainv} with respect to $z^*$ gives
\begin{equation}
\label{eq:dStarsC}
\partial_z^{*} \big[\tilde{\bm{\Sigma}}_{j}^{-1}\big]
= i \bm{\sigma}_{-}
+\sum_{\tau \in \partial j}
\bm R_{x_{j\tau}y_{j\tau}}\!\big(\partial_z^{*} \bm W_{\bm \Lambda_\tau^{(j)}}\big),
\end{equation}
and using $\partial_z^{*}\tilde{\bm{\Sigma}}_{j}= - \tilde{\bm{\Sigma}}_{j}\,[\partial_z^{*}\tilde{\bm{\Sigma}}^{-1}_{j}]\,\tilde{\bm{\Sigma}}_{j}$ we can evaluate \(\rho_{\bm F}\).
The derivatives $\partial_z^{*} \bm W_{\bm \Lambda_\tau^{(j)}}$ are obtained by differentiating
\eqref{eq:deltaCavity}–\eqref{eq:WCavity}, leading to
\begin{equation}
\label{eq:dStarCavityEquations}
\begin{split}
\partial_z^{*}\bm \Lambda^{(j)}_{t}&=
\frac{1}{d}\sum_{i \in \partial t \setminus j}
\begin{pmatrix}
  x_{it}^{2}\partial_z^{*}\tilde{\bm{\Sigma}}^{(t)}_{i} &
  y_{it} x_{it}\partial_z^{*}\tilde{\bm{\Sigma}}^{(t)}_{i} \\[0.4em]
  y_{it} x_{it}\partial_z^{*}\tilde{\bm{\Sigma}}^{(t)}_{i} &
  y_{it}^{2}\partial_z^{*}\tilde{\bm{\Sigma}}^{(t)}_{i}
\end{pmatrix},
\\[4pt]
\partial_z^{*} \tilde{\bm{\Sigma}}^{(t)}_{j}
&= 
-\tilde{\bm{\Sigma}}^{(t)}_{j}\Bigl(
i \bm{\sigma}_{-}
+\sum_{\tau \in \partial j \setminus t}
\bm R_{x_{j\tau}y_{j\tau}}\!\big(\partial_z^{*} \bm W_{\bm \Lambda_\tau^{(j)}}\big)
\Bigr)\tilde{\bm{\Sigma}}^{(t)}_{j},
\\[4pt]
\partial_z^{*}\bm W_{\bm \Lambda_\tau^{(j)}}
&=
-\,\bm W_{\bm \Lambda_\tau^{(j)}} \, \partial_z^{*}\bm{\Lambda}_{\tau}^{(j)} \, \bm W_{\bm \Lambda_\tau^{(j)}}.
\end{split}
\end{equation}
This companion system can be solved numerically via the same fixed-point strategy.

\section{Computing the average spectral density using the replica method}
\label{app:asymmetricReplica}
In this appendix we compute the average spectral density using the replica method. Our starting point is to compute the average over the ensembles of matrices $\bm F$ of the spectral density as written in equation (\ref{eq:rhoSpinors0}). To compute the average of the logarithm, we use the \emph{replica trick}, a procedure based on the following mathematical identity,
\begin{equation}
  \begin{split}
\exValueF{\ln Z}= \lim_{n \rightarrow0} \frac{1}{n}\ln \exValueF{Z^n},
  \end{split}
\end{equation}
where $\exValueF{\cdot}$ denotes the average over the ensemble of matrices $\bm F$. The idea of the replica trick is to compute $\frac{1}{n}\ln \exValueF{Z^n}$ for an arbitrary positive integer $n$, and then to continue analytically the result to $n=0$. Even though the analytical continuation of a function defined over the integers in not unique, it is well-known that the one obtained from the \emph{replica symmetric ansatz} gives the correct result for models on Poisson graphs \cite{Mezard1986_2}, so this is the approach we will pursue.

To find $\exValueF{Z^n}$ for integer $n$, we compute the partition function of an augmented system made by $n$ copies --- or replicas --- of the original system. Before going on, we dwelve a little bit in the notation we will using for this appendix. By $\psi_{i}^{a}$, with $i=1, \dots,N$ and $a=1, \dots,n$, we denote the spinor located at node $i$ for the $a$-th replica. From this point on, the indices $i,j,$ take values inside the range $1, \dots, N$; $a,b$, inside $1, \dots,n$; and $t,\tau$ inside $1, \dots,T$. An index between brackets, indicates that such index is not fixed, but rather, it is allowed to vary along all its possible values. For instance, $\psi_{1}^{\{a\}}$ represents the tuple $(\psi_{1}^{1},\psi_{1}^{2}, \dots,\psi_{1}^{n})$, while $\psi_{\{i\}}^{1}$ represents the tuple $(\psi_{1}^{1},\psi_{2}^{1}, \dots,\psi_{N}^{1})$. With this notation, and using (\ref{eq:HFpsixy}) we write,
\begin{equation}
  \begin{split}
\exValueF{Z^n}&=
\int d \psi_{\{i\}}^{\{a\}}
e^{-\sum_{i,a} \psi^a_{i}{}^{\dagger}{\bm{M}} \psi^a_{i}}
    \exValuebF{
e^{
   -i\sum_{a,t}  \mathxy{Z}^a_{t}{}^{\dagger}
    \Smat
\mathxy{Z}^a_{t}
  }
    }
   \\&=
\int d \psi_{\{i\}}^{\{a\}}
e^{-\sum_{i,a} \psi^a_{i}{}^{\dagger}{\bm{M}} \psi^a_{i} }
    \left[\exValuebF{
e^{-i \sum_{a} \mathxy{Z}^a_{1}{}^{\dagger}
    \Smat
\mathxy{Z}^a_{1}
    }
    }\right]^T
  \end{split}
\end{equation}
where $d \psi_{\{i\}}^{\{a\}}$ indicates an integral over all the spinors with their replicas, we defined $\mathxy{Z}^a_t=(\mathxy{X}^a_t,\mathxy{Y}^a_t)$, and we used the fact that nodes with different $t$ are independent by definition to obtain the last equation.

The next step, is to compute the average
$\exValuebF{
e^{-i \sum_{a} \mathxy{Z}^a_{1}{}^{\dagger}
    \Smat
\mathxy{Z}^a_{1}
    }
    }$.
Consider the subspace $\Omega_k$ of matrices $\bm F$ where exactly $k$ of the nodes $i$ are connected with the node $t=1$. In the thermodynamic limit, the probability that a sampled matrix $\bm F$  is inside $\Omega_k$ is,
\begin{equation}
  \begin{split}
p_k=\binom{N}{k} \left(\frac{d}{N}\right)^k\left(1-\frac{d}{N}\right)^{N-k} \approx
\frac{d^k e^{-d}}{k!},
  \end{split}
  \label{eq:pkDef}
\end{equation}
where we used Stirling's approximation for the last expression. Notice that, the average we are interested in, only depends on $\mathxy{Z}^{\{a\}}_1$. When restricted to $\Omega_k$, $\mathxy{Z}^{\{a\}}_1$, can be sampled in the following way,
\begin{itemize}
  \item Sample uniformly $k$ nodes $i_{\lambda}$, $\lambda= 1, \dots ,k$ from the set of nodes $1, \dots, N$, avoiding repetitions of the indices $i_\lambda$.
  These represent the nodes that are connected to the node $t=1$.
  \item Sample independently  $k$ pairs $(x_\lambda, y_ \lambda)$ from $\rho$.
  \item Compute ${\mathxy Z}^{\{a\}}_{1}=d^{-1}\sum_{\lambda =1}^{k}( x_{\lambda} \psi^{\{a\}}_{i_\lambda}, y_{\lambda} \psi^{\{a\}}_{i_\lambda})$.
\end{itemize}
In the thermodynamic limit, repetitions among the indices $i_\lambda$ are avoided
almost surely, so the first step is equivalent to sample $k$ spinors $\psi^{\{a\}}$
independent and identically from the following probability distribution,
\begin{equation}
  \begin{split}
P(\psi^{\{a\}}) \equiv \frac{1}{N}\sum_{i=1}^N \delta ( \psi^{\{a\}}- \psi_i^{\{a\}}).
  \end{split}
  \label{eq:PEmpiricalDef}
\end{equation}
Although $P$ depends on $\psi_{\{i\}}^{\{a\}}$, we do not write this dependence explicitly to avoid cluttering the notation. By the previous analysis we conclude (here, the values allowed for $\lambda$ are $1, \dots,k$),
\begin{equation*}
  \begin{split}
    \exValuebF{
    e^{
   -i \sum_{a}  {\mathxy Z}^a_{1}{}^{\dagger}
    \Smat
{\mathxy Z}^a_1
    }
    }
    &=
\sum_{k=0}^\infty p_k \int d \psi^{\{a\}}_{\{{\lambda}\}} d x_{\{\lambda\}} d y_{\{\lambda\}}
\left[ \prod_{\lambda=1}^k P(\psi^{\{a\}}_{\lambda}) \rho(x_\lambda,y_\lambda) \right]
e^{
   -i \sum_{a}  {\mathxy Z}^a_{1;k}{}^{\dagger}
    \Smat
   {\mathxy Z}^a_{1;k}
    }
\\
    &=
\sum_{k=0}^\infty p_k \int d \psi^{\{a\}}_{\{{\lambda}\}} d x_{\{\lambda\}} d y_{\{\lambda\}} d {Z}^{\{a\}}
\left[ \prod_{\lambda=1}^k P(\psi^{\{a\}}_{\lambda}) \rho(x_\lambda,y_\lambda) \right]
\\&\qquad\qquad\delta ({Z}^{\{a\}}_{} -{\mathxy Z}^{\{a\}}_{1;k} )
 e^{
   -i \sum_{a}
   {Z}^a{}^\dagger
    \Smat
   {Z}^a
    }
\\& \equiv
    \int d{Z}^{\{a\}} \mathcal M_P({Z}^{\{a\}})
e^{
   -i \sum_{a}  {Z}^a{}^{\dagger}
    \Smat
   {Z}^a
    }
  \end{split}
\end{equation*}
where we defined
\begin{equation}
  \begin{split}
{\mathxy Z}^{\{a\}}_{1;k}=\frac{1}{d}\sum_{\lambda =1}^{k}( x_{\lambda} \psi^{\{a\}}_{\lambda},y_{\lambda}\psi^{\{a\}}_{\lambda}),
  \end{split}
  \label{eq:Z_aik}
\end{equation}
to introduce a Dirac delta over the space of spinors $Z^{\{a\}}$; and $\mathcal M_P({Z}^{\{a\}})$ denotes the following probability density for $Z^{\{a\}}$,
\begin{equation}
  \begin{split}
\mathcal M_P({Z}^{\{a\}})&=
\sum_{k=0}^\infty p_k \int d \psi^{\{a\}}_{\{{\lambda}\}} d x_{\{\lambda\}} d y_{\{\lambda\}}
 \left[\prod_{\lambda=1}^k P(\psi^{\{a\}}_{\lambda}) \rho(x_\lambda,y_\lambda)\right]
\delta ({Z}^{\{a\}}_{} - {\mathxy Z}^{\{a\}}_{1;k} )\,.
  \end{split}
  \label{eq:MDef}
\end{equation}
Note that ${\mathxy Z}^{\{a\}}_{1;k}$ is a function of $\psi^{\{a\}}_{\{\lambda\}}$, while ${Z}^{\{a\}}_{}$ denotes an integration variable. From this analysis we conclude,
\begin{equation}
  \begin{split}
\exValueF{Z^n}&=
\int d \psi_{\{i\}}^{\{a\}}
e^{-\sum_{i,a} \psi^a_{i}{}^{\dagger}{\bm{M}} \psi^a_{i}}
    \left[\int d {Z}^{\{a\}} \mathcal M_P({Z}^{\{a\}})
e^{
   -i \sum_{a}  {Z}^a_{}{}^{\dagger}
    \Smat
   {Z}^a_{}
    }
\right]^T,
  \end{split}
\end{equation}
To compute the integral over $\psi^{\{a\}}_{\{i\}}$, we introduce a Dirac delta over the space of probabilities $\tilde P(\psi^{\{a\}})$ ---the space where $P$ from Eq.~(\ref{eq:PEmpiricalDef}) lives. In the thermodynamic limit, this Dirac delta tends to a functional Dirac delta, but we denote it as an ordinary Dirac delta to simplify the notation. The result is,
\begin{equation*}
  \begin{split}
\exValueF{Z^n}&=
\int d \psi_{\{i\}}^{\{a\}} d\tilde P
\delta(P-\tilde P)
e^{ -\sum_{i,a} \psi^a_{i}{}^{\dagger}{\bm{M}} \psi^a_{i}}
    \left[\int d {Z}^{\{a\}} \mathcal M_{\tilde P}({Z}^{\{a\}})
e^{
   -i \sum_{a}  {Z}^a_{}{}^{\dagger}
    \Smat
   Z^a_{}
    }
\right]^T.
  \end{split}
\end{equation*}
The next step is to write $\delta (\tilde P-P)$ in the Fourier representation, 
\begin{equation}
  \begin{split}
\delta (\tilde P-P)&= \int d \hat P e^{i N \int d \psi^{\{a\}}\hat P(\psi^{\{a\}}) [\tilde P(\psi^{\{a\}})-P(\psi^{\{a\}})]}
\\&=
\int d \hat P e^{i N \int d \psi^{\{a\}} \hat P(\psi^{\{a\}}) \tilde P(\psi^{\{a\}})} e^{-i   \sum_j \hat P(\psi_j^{\{a\}}) },
  \end{split}
\end{equation}
where the integral $d \hat P$ is over the space of all functions of a spinor $\psi^{\{a\}}$, and we used (\ref{eq:PEmpiricalDef}) for the last expression; all the missing normalization factors are absorbed in $d \hat P$. Within this framework, we can write,
\begin{equation}
  \begin{split}
\exValueF{Z^n} = \int d \hat P d \tilde P e^{-N \mathrm{S} (\hat P,\tilde P)},
  \end{split}
  \label{eq:defActionS}
\end{equation}
where we defined the following \emph{action},
\begin{equation}
  \begin{split}
\mathrm{S}(\hat P,\tilde P)=&
    -i  \int d \psi^{\{a\}} \hat P(\psi^{\{a\}}) \tilde P(\psi^{\{a\}})
    -\frac{1}{q}\ln\left[\int d {Z}^{\{a\}}\mathcal M_{\tilde P}({Z}^{\{a\}})
e^{
   -i \sum_{a}  {Z}^a_{}{}^{\dagger}
    \Smat
   {Z}^a_{}
    }
\right]
\\&-
\ln\left[\int d \psi^{\{a\}} e^{-i  \hat P(\psi^{\{a\}}) } e^{-\sum_{a} \psi^a{}^{\dagger}{\bm{M}} \psi^a}\right],
  \end{split}
  \label{eq:freeEnergy0}
\end{equation}
and we remind the reader that $q=N/T$. In the thermodynamic limit, we can compute the integral using the saddle point method. The saddle equations are,
\begin{equation}
  \begin{split}
    0=
\frac{\delta \mathrm{S}}{\delta\hat P(\psi^{\{a\}})}&=
    -i \tilde P (\psi^{\{a\}}) + i \frac{e^{-i  \hat P(\psi^{\{a\}}) } e^{-\sum_{a} \psi^a{}^{\dagger}{\bm{M}} \psi^a}}{\int d \psi^{\{b\}} e^{-i  \hat P(\psi^{\{b\}}) } e^{-\sum_{b} \psi^b{}^{\dagger}{\bm{M}} \psi^b}},
  \label{eq:saddlesaddle1}
\\
    0=
\frac{\delta \mathrm{S}}{\delta\tilde P(\psi^{\{a\}})}&=
    -i   \hat P(\psi^{\{a\}}) -
\frac{\int d {Z}^{\{a\}}\frac{\delta \mathcal M_{\tilde P}({Z}^{\{a\}})}{\delta \tilde P (\psi^{\{a\}})\vphantom{A^{A^a}}}
e^{
   -i \sum_{a}  {Z}^a{}^{\dagger}
    \Smat
    {Z}^a_{}
    }
}
{q \int d {Z}^{\{a\}} \mathcal M_{\tilde P}({Z}^{\{a\}})
e^{
   -i \sum_{a}  {Z}^a_{}{}^{\dagger}
    \Smat
   {Z}^a_{}
    }
},
  \end{split}
\end{equation}
where, an explicit computation using (\ref{eq:Z_aik}), (\ref{eq:MDef}) reveals,
\begin{equation}
  \begin{split}
\frac{\delta \mathcal{M}_{\tilde P}({Z}^{\{a\}})}{\delta  \tilde P ( \psi^{\{a\}})}&=
d\int  dx d y\,  \rho(x,y)
\mathcal{M}_{\tilde P}\left({Z}^{\{a\}}- \frac{1}{\sqrt d} (x \psi^{\{a\}}, y \psi^{\{a\}})\right).
  \end{split}
  \label{eq:dMdP}
\end{equation}
The saddle equations (\ref{eq:saddlesaddle1}) are challenging to solve, even numerically. Further advances can be made by assuming that $\tilde P$ and $\hat P$ are a superposition of Gaussians where the entries of $\psi^{\{a\}}$ are independent and identically generated. Specifically, we propose the following replico-symmetric ansatz,
\begin{equation}
  \begin{split}
\tilde P ( \psi^{\{a\}})&= \int d\tilde {\bm{\Sigma}}\,  \tilde \omega (\tilde {\bm{\Sigma}}) \prod_a \frac{e^{- \psi^a{}^\dagger \tilde {\bm{\Sigma}}^{-1} \psi^a }}{\det \tilde {\bm{\Sigma}}},\quad
\\
\hat P ( \psi^{\{a\}})&= \mathcal A \int d \hat {\bm{\Sigma}}\,  \hat \omega ( \hat {\bm{\Sigma}}) \prod_a \frac{e^{- \psi^a{}^\dagger  \hat {\bm{\Sigma}}^{-1} \psi^a }}{\det  \hat {\bm{\Sigma}}},\quad
    \\
\mathcal M_{\tilde P}( {Z}^{{\{a\}}}) &=  \int d {\bm{\Lambda}}\, m( {\bm{\Lambda}}) \prod_a \frac{   e^{-  {Z}^a{}^{\dagger}
    {\bm{\Lambda}}^{-1}
   Z^a}
}{\det {\bm{\Lambda}} }
  \end{split}
  \label{eq:rsAnsatzAll}
\end{equation}
where $\tilde{\omega}$ and $\hat{\omega}$ denote measures on $\mathbb{C}^{2 \times 2}$, and $m$ denotes a measure on $\mathbb{C}^{4 \times 4}$. Mathematically, they represent the weights of the various Gaussian components in the ansatz. Here $\mathcal A$ is constant equal to $\mathcal A= \int d \psi^{\{a\}}\hat P ( \psi^{\{a\}})$ reflecting that $\hat P$ is in general, not normalized to one. Direct substitution of this ansatz in (\ref{eq:saddlesaddle1}) produces a relation between $\tilde \omega$, $\hat \omega$ and $m$.

The motivation to consider such an ansatz it at least three fold. Firstly, since Gaussian distributions are closed under several algebraic operations, we can use it in (\ref{eq:saddlesaddle1}) and obtain a consistent system of equations. Secondly, the Gaussian distributions involved are symmetric in the replica index $a$, indicating that the solution is symmetric under the exchange of replicas, and in this problem we do not expect any replica symmetry breaking to take place \cite{Mezard1986_2}. And finally, it allow us to write $\hat P$, $\tilde P$ and $\mathcal {M}$ in terms of the probability densities $\hat \omega$, $\tilde \omega$ and $m$ defined over $\mathbb{C}^{2 \times 2}$ and $\mathbb{C}^{4 \times 4}$. Since these spaces are independent of $n$, the limits when $n$ approaches zero of $\hat \omega$, $\tilde \omega$ and $m$ are probability densities defined over them. These limits are simpler to interpret and to work with than the respective limits of $P$, $\tilde P$ and $m$.

We proceed to find the equations to determine $\tilde \omega$, $\hat \omega$, $m$.  After using the replico-symmetric ansatz (\ref{eq:rsAnsatzAll}), we can write the second equation  of (\ref{eq:saddlesaddle1}) in terms of $m$. By comparing this result with the ansatz we conclude,
\begin{equation}
  \begin{split}
 \hat \omega (\hat {\bm{\Sigma}})=
\frac{id }{q \mathcal A\mathcal M_0}
 \int d {\bm{\Lambda}}\, m( {\bm{\Lambda}})
\exValueb{\delta\left(\hat {\bm{\Sigma}}-
\left\{ {\bm{R}}_{xy}[
    {\bm{W}}_{\bm{\Lambda}}
]\right\}^{-1}\right)}_{\rho}
\frac{ \det^n( \hat {\bm{\Sigma}})}{\det^n (\bm{1}_4+i {\bm{\Lambda}} \Smat )},
  \end{split}
  \label{eq:finalSaddle2}
\end{equation}
where $\bm R$ and $\bm W$ are precisely the ones defined in Eqs.~(\ref{eq:RMainDef})--(\ref{eq:WCavity0}), $\exValueb{\cdot}_\rho$ denotes the average computed from the distribution $\rho$, and,
\begin{equation}
  \begin{split}
\mathcal{M}_0= \int d {Z}^{\{a\}} \mathcal M_{\tilde P({Z}^{\{a\}})}
e^{-i \sum_{a}  {Z}^a{}^{\dagger}\Smat {Z}^a}
    =
\int d {\bm{\Lambda}}\, m( {\bm{\Lambda}})
\frac{1}{\det^n (\bm{1}_4+i {\bm{\Lambda}}\Smat) }.
  \end{split}
  \label{eq:hereIsMo}
\end{equation}
Next, we compute  $\mathcal {M}$  in terms of $\tilde \omega$. After
using (\ref{eq:rsAnsatzAll}) in (\ref{eq:MDef}) 
and comparing with (\ref{eq:rsAnsatzAll})
it follows that,
\begin{equation}
  \begin{split}
m({\bm{\Lambda}})&=
                                                          \sum_{k=0}^\infty p_k
\int d\tilde {\bm{\Sigma}}_{{\{\lambda\}}} \, \left[\prod_{\lambda=1}^k \tilde \omega (\tilde {\bm{\Sigma}}_\lambda)\right]
 \exValueb{  \delta \left( {\bm{\Lambda}}-
\frac{1}{d}\sum\limits_\lambda
\begin{pmatrix}
 x_\lambda^2 \tilde {\bm{\Sigma}}_\lambda & x_\lambda y_\lambda \tilde {\bm{\Sigma}}_\lambda \\
  x_\lambda y_\lambda\tilde {\bm{\Sigma}}_\lambda & y_\lambda^2\tilde  {\bm{\Sigma}}_\lambda \\
\end{pmatrix}
\right)
    }_{\rho^k}
             \\ & \times
\frac{ \det^n( {\bm{\Lambda}})}{\det^n ( \bm{1}_4+i {\bm{\Lambda}} \Smat)}
  \end{split}
  \label{eq:finalSaddle3}
\end{equation}
where the pairs $(x_\lambda,y_\lambda)$, $\lambda=1, \dots,k$ are sampled independently and identically from $\rho$ to compute $\exValueb{\cdot}_{\rho^k}$.

Finally,  after using the replico-symmetric ansatz, and doing a Taylor expansion of the exponential term of the first saddle equation (\ref{eq:saddlesaddle1}) we conclude,
\begin{equation}
  \begin{split}
\tilde \omega(\tilde {\bm{\Sigma}})=
\frac{1}{\mathcal N}
\sum_{k=0}^\infty \frac{(-i \mathcal A)^k}{k!}
\int d \hat {\bm{\Sigma}}_{{\{\lambda\}}} \left[\prod_{\lambda=1}^k \hat \omega(\hat {\bm{\Sigma}}_\lambda)\right]
 \delta\Big(\tilde {\bm{\Sigma}}-\Big[\sum_\lambda \hat {\bm{\Sigma}}_\lambda^{-1}+{\bm{M}}\Big]^{-1}\Big)
\frac{\det^n (\tilde {\bm{\Sigma}})}{ \prod_\lambda\det^{n} (\hat {\bm{\Sigma}}_\lambda)}
  \end{split}
  \label{eq:almostNoLabel}
\end{equation}
where $\mathcal N$ is a normalization factor.

The system of equations (\ref{eq:finalSaddle3}), (\ref{eq:finalSaddle2}) and (\ref{eq:almostNoLabel}) can be solved numerically for arbitrary $n$ using a weighted \emph{population dynamics} algorithm \cite{Metz_2016}. In particular, they can be solved for $n=0$. In this case, from (\ref{eq:hereIsMo}) we conclude $\mathcal M_0=1$.
Using the fact that the integral $\hat \omega$ and $\tilde \omega$ over the respective spaces is one, we also conclude $\mathcal A=i d/q$ and $\mathcal N=e^{d/q}$.

Thus the equations simplify to,
 \begin{equation}
  \begin{split}
 \hat \omega (\hat {\bm{\Sigma}})&=
 \int d {\bm{\Lambda}}\, m( {\bm{\Lambda}})
\exValueb{\delta\left(\hat {\bm{\Sigma}}-
\left\{ {\bm{R}}_{xy}[
    {\bm{W}}_{\bm{\Lambda}}
]\right\}^{-1}\right)}_{\rho},
                  \\
\tilde \omega(\tilde {\bm{\Sigma}})&= e^{-d/q}
\sum_{k=0}^\infty \frac{d^k}{q^k k!}
\int d \hat {\bm{\Sigma}}_{{\{\lambda\}}} \left[\prod_{\lambda=1}^k \hat \omega(\hat {\bm{\Sigma}}_\lambda)\right]
 \delta\Big(\tilde {\bm{\Sigma}}-\Big[\sum_\lambda \hat {\bm{\Sigma}}_\lambda^{-1}+{\bm{M}}\Big]^{-1}\Big),
\\
m({\bm{\Lambda}})&=
                                                          \sum_{k=0}^\infty p_k
\int d\tilde {\bm{\Sigma}}_{{\{\lambda\}}} \, \left[\prod_{\lambda=1}^k \tilde \omega (\tilde {\bm{\Sigma}}_\lambda)\right]
 \exValueb{  \delta \left( {\bm{\Lambda}}-
\frac{1}{d}\sum\limits_\lambda
\begin{pmatrix}
 x_\lambda^2 \tilde {\bm{\Sigma}}_\lambda & x_\lambda y_\lambda \tilde {\bm{\Sigma}}_\lambda \\
  x_\lambda y_\lambda\tilde {\bm{\Sigma}}_\lambda & y_\lambda^2\tilde  {\bm{\Sigma}}_\lambda \\
\end{pmatrix}
\right)
    }_{\rho^k}
  \end{split}
  \label{eq:saddleFamily}
\end{equation}
The resulting probability densities  from solving (\ref{eq:saddleFamily}) can be used to compute the spectral density. By making a Taylor  expansion of (\ref{eq:freeEnergy0}) for small $n$ we conclude,
\begin{equation*}
  \begin{split}
\mathrm{S}(\hat P,\tilde P)=&
     n
\Bigg\{
i\int d\hat {\bm{\Sigma}} d\tilde {\bm{\Sigma}} \,\hat \omega(\hat {\bm{\Sigma}}) \tilde \omega(\tilde {\bm{\Sigma}}) \ln\det (\hat {\bm{\Sigma}}+\tilde {\bm{\Sigma}})
+\frac{1}{q}
\int d {\bm{\Lambda}}\, m( {\bm{\Lambda}}) \det (\bm{1}_4+i{\bm{\Lambda}} \Smat)
           \\& 
    +
e^{-d/q}
\sum_{k=0}^\infty 
\frac{d^k}{q^k k!}
\int d\hat {\bm{\Sigma}}_{{\{\lambda\}}} \left[\prod_{\lambda=1}^k \hat \omega(\hat {\bm{\Sigma}}_\lambda)\right]
    \ln\Big(\det\Big[\sum\limits_\lambda\hat  {\bm{\Sigma}}_\lambda^{-1}+{\bm{M}}\Big]\Big)
           \\&
+                  
\frac{d}{q}
\int d\hat {\bm{\Sigma}}\, \hat \omega(\hat {\bm{\Sigma}}) \ln\left(\det \hat {\bm{\Sigma}}\right)
\Bigg\}
    +\mathcal O(n^2).
  \end{split}
\end{equation*}
From the previous equation, it is straightforward to compute $\exValueF{\ln Z} = \lim_{n \rightarrow 0} \frac{1}{n} \ln \exValueF{Z^n}$ using the saddle point method and equation (\ref{eq:defActionS}). To compute the spectral density (\ref{eq:rhoSpinors0}), we need to compute the derivative $\partial_z\exValueF{\ln Z}$.

While we did not write it explicitly, $\tilde \omega$, $\hat \omega$ and $m$ depend on $z$, since ${\bm{M}}$ depends on $z$. However, since they were obtained by an extremization procedure, when computing the first derivative $\partial_z$, such dependence can be ignored \cite{Nishimori_2001}. Thus,
\begin{equation}
  \begin{split}
  \partial_z \exValueF{\ln Z}&=-N
e^{-d/q}
\sum_{k=0}^\infty
\frac{d^k}{q^k k!}
\int d\hat {\bm{\Sigma}}_{{\{\lambda\}}} \prod_{\lambda=1}^k \hat \omega(\hat {\bm{\Sigma}}_\lambda)\,
\mathrm{tr}\Big\{\Big[\sum\limits_\lambda \hat {\bm{\Sigma}}_\lambda^{-1}+{\bm{M}}\Big]^{-1} \partial_z {\bm{M}}\Big\}
     \\&=-i N
    \int d \tilde {\bm{\Sigma}}\,
\tilde{\omega}(\tilde {\bm{\Sigma}})\,\mathrm{tr}\big[\tilde {\bm{\Sigma}} \bm{\sigma}_+\big]
     =-i N
    \int d \tilde {\bm{\Sigma}}\,
\tilde{\omega}(\tilde {\bm{\Sigma}})\tilde {{\Sigma}}_{21} ,
  \end{split}
\end{equation}
where we used the saddle equation (\ref{eq:saddleFamily}) to obtain the second line.
Therefore, the average spectral density can be computed as follows,
\begin{equation}
  \begin{split}
\exValueF{\rho(z)}=
\frac{i}{\pi}\partial_z^{*}
    \int d \tilde {\bm{\Sigma}}\,
\tilde{\omega}(\tilde {\bm{\Sigma}})\tilde {{\Sigma}}_{21}\,.
  \end{split}
\end{equation}
To compute the derivative $\partial_z^*$ of the previous equation, the omitted dependence of $\tilde \omega$ on $z^*$ has to be considered. This can be done by introducing some new variables $\hat {\bm{\Sigma}}^\star=\partial_z^*\hat {\bm{\Sigma}}$, $\tilde {\bm{\Sigma}}^\star=\partial_z^*\tilde {\bm{\Sigma}} $, $\hat {\bm{\Lambda}}^\star=\partial_z^*\hat {\bm{\Lambda}}$, and computing the derivatives of
(\ref{eq:saddleFamily}) to find the joint probabilities $\tilde \omega(\tilde {\bm{\Sigma}}, \tilde {\bm{\Sigma}}^\star)$, $\hat \omega(\hat {\bm{\Sigma}}, \hat {\bm{\Sigma}}^\star)$ and $m(\hat {\bm{\Lambda}},\hat {\bm{\Lambda}}^\star)$. The expressions can be further simplified if we write them in terms of  $\hat {\bm{\Gamma}}= \hat {\bm{\Sigma}}^{-1}$, $\hat {\bm{\Gamma}}^\star= \partial_z^* \hat {\bm{\Gamma}}$. The result for the joint probabilities is the one in Eq.~(\ref{eq:saddleCompleteFamily}).

The spectral density is simply,
\begin{equation}
  \begin{split}
\exValueF{\rho(z)}=
\frac{i}{\pi}
    \int d \tilde {\bm{\Sigma}} d  \tilde {\bm{\Sigma}}^\star\, \tilde{\omega}(\tilde {\bm{\Sigma}}, \tilde {\bm{\Sigma}}^\star) \tilde {{\Sigma}}^\star_{21}\,.
  \end{split}
\end{equation}

\section{Equivalence between the replica and cavity methods}
\label{app:replicaCavity}
In the previous appendices we derived the spectral density using both the cavity and replica methods. The cavity method is conceptually and algebraically simpler, while the replica method is numerically convenient because it works directly in the thermodynamic limit. In this section we show that the two approaches are equivalent. This is a special case of a general result for Poisson graphs \cite{Mezard2001}.

In the previous appendix we showed that the average spectral density can be expressed in
terms of the probability densities $\hat \omega$ and $\tilde \omega$, obtained
from the replica–symmetric ansatz in the limit $n\to 0$. Here we show that $\hat \omega$ and $\tilde \omega$ are the probability densities arising from averaging the cavity quantities, providing a simpler interpretation of these measures.

The key observation is that $\tilde {\bm{\Sigma}}^{(t)}_j$ and ${\bm{\Delta}}_{t}^{(j)}$ depend on the matrix $\bm F$, so they are random variables with associated probability distributions.

Consider all matrices $\bm F$ in the ensemble such that the nodes $t$ and $j$ of the underlying graph are connected. Because we condition on the presence of this edge,
the matrices $\tilde {\bm{\Sigma}}^{(t)}_j$ and ${\bm{\Delta}}_{t}^{(j)}$ are well-defined. For ease of comparison with the replica method, also define
\begin{equation}
  \begin{split}
\hat {\bm{\Sigma}}_{t}^{(j)} =  [{\bm{R}}_{x_{j t} y_{j t}} ({\bm{W}}_{{\bm{\Lambda}}^{(j)}_t})]^{-1} \, .
  \end{split}
  \label{eq:tildeSigmaCavityToReplica}
\end{equation}
With a modest amount of foresight, denote by $\tilde \omega$ the probability density of $\tilde {\bm{\Sigma}}^{(t)}_j$, and by $\hat \omega$ that of $\hat {\bm{\Sigma}}^{(t)}_j$. Since the results are independent of the particular $t$ and $j$, we omit these indices in $\hat \omega$ and $\tilde \omega$.

The first cavity equation (\ref{eq:cavEquationsMain}) can be written as
\begin{equation}
  \begin{split}
(\tilde {\bm{\Sigma}}^{(t)}_j{})^{-1} = {\bm{M}} + \sum_{\tau \in \partial j\setminus t} (\hat {\bm{\Sigma}}_{\tau}^{(j)})^{-1} \, .
  \end{split}
\end{equation}
Let $\tilde \Omega_k$ be the set of graphs in which $j$ has $k$ additional neighbors—besides $t$—denoted $\tau_1,\dots,\tau_k$. Since the presence or absence of each edge is independent of all other edges, in the thermodynamic limit, the probability $\tilde p_k$ that the graph of a matrix $\bm F$ lies in $\tilde \Omega_k$ is obtained analogously to $p_k$ from Eq.~(\ref{eq:pkDef}), with the only modification that the number of nodes $\tau$ potentially connected to $j$ is $T=N/q$ rather than $N$. Thus,
\begin{equation}
\tilde p_k= e^{-d/q}\frac{d^k}{q^k k!}\,.
\end{equation}

Within $\tilde \Omega_k$,
\begin{equation}
  \begin{split}
(\tilde {\bm{\Sigma}}^{(t)}_j{})^{-1} = {\bm{M}} + \sum_{\lambda=1}^k (\hat {\bm{\Sigma}}_{\tau_\lambda}^{(j)})^{-1} \, .
  \end{split}
\end{equation}
In this case, $\tilde {\bm{\Sigma}}^{(t)}_j$ depends only on the random variables $\hat {\bm{\Sigma}}_{\tau_\lambda}^{(j)}$.
For the system with node $j$ removed these variables are statistically independent and identically distributed.
Therefore, letting $\tilde \omega_k$ denote the probability density of $\tilde {\bm{\Sigma}}^{(t)}_j$ conditional on $k$ additional neighbors, we have
\begin{equation}
  \begin{split}
    \tilde \omega_k (\tilde {\bm{\Sigma}}^{(t)}_j{}  )=
 \int \Bigg[\prod_{\lambda = 1}^k  \hat \omega(\hat {\bm{\Sigma}}_{\lambda})d \hat {\bm{\Sigma}}_\lambda\Bigg] \delta\left( \tilde {\bm{\Sigma}}^{(t)}_j - \Bigg({\bm{M}} + \sum_{\lambda=1}^k \hat {\bm{\Sigma}}_{\lambda}^{-1}\Bigg)^{-1}\right),
  \end{split}
\end{equation}
and the full probability density is
\begin{equation}
  \begin{split}
    \tilde \omega(\tilde {\bm{\Sigma}}^{(t)}_j{} )= \sum_{k=0}^\infty \tilde{p}_k \int \Bigg[\prod_{\lambda = 1}^k \hat \omega(\hat {\bm{\Sigma}}_{\lambda}) d \hat {\bm{\Sigma}}_\lambda\Bigg]
    \delta\left( \tilde {\bm{\Sigma}}^{(t)}_j - \Bigg({\bm{M}} + \sum_{\lambda=1}^k \hat {\bm{\Sigma}}_{\lambda}^{-1}\Bigg)\right).
  \end{split}
\end{equation}
This is the second equation obtained from the replica method.

A second equation follows from (\ref{eq:tildeSigmaCavityToReplica}). In the system where $j$ is removed, $ {\bm{\Lambda}}^{(j)}_t$ and the pair $(x_{jt},y_{jt})$ are independent, since $ {\bm{\Lambda}}^{(j)}_t$ depends only on $(x_{it},y_{it})$ with $i \in \partial j \setminus t$. Therefore, the density of $\hat {\bm{\Sigma}}^{(t)}_j$ is
\begin{equation}
  \begin{split}
\hat \omega(\hat {\bm{\Sigma}}^{(t)}_j)= \int dx\,dy\, d {\bm{\Lambda}}\, m({\bm{\Lambda}})\, \rho(x, y)\,
 \delta\left(  \hat {\bm{\Sigma}}^{(t)}_j- \{{\bm{R}}_{x y} ({\bm{W}}_{\bm{\Lambda}})\}^{-1}\right).
  \end{split}
\end{equation}
Similarly, from the second of (\ref{eq:cavEquationsMain}), ${\bm{\Lambda}}_{t}^{(j)}$ depends only on $\tilde{\bm{\Sigma}}_{t}^{(i)}$ and $(x_{it},y_{it})$ with $i \in \partial t\setminus j$. In the system where $j$ is removed these variables are all independent. The probability that $t$ has $k$ additional neighbors is again $p_k$. Combining the arguments used above yields
\begin{equation}
  \begin{split}
m ({\bm{\Lambda}}_{t}^{(j)}) &=
\sum_{k=0}^\infty p_k  \int \Bigg[\prod_{\lambda=1}^k dx_\lambda\,dy_{\lambda}\,d \tilde {\bm{\Sigma}}_\lambda \;\rho(x_\lambda, y_\lambda)\, \tilde \omega(\tilde{\bm{\Sigma}}_\lambda )\Bigg]
                         \\& \times
 \delta \left( {\bm{\Lambda}}-
\left[\frac{1}{d}\sum_{\lambda=1}^k
\begin{pmatrix}
 x_\lambda^2 \tilde {\bm{\Sigma}}_\lambda & x_\lambda y_\lambda \tilde {\bm{\Sigma}}_\lambda \\
  x_\lambda y_\lambda\tilde {\bm{\Sigma}}_\lambda & y_\lambda^2\tilde  {\bm{\Sigma}}_\lambda \\
\end{pmatrix}
\right]\right).
  \end{split}
\end{equation}
This is the last equation obtained from the replica formalism.

Finally, note that in the thermodynamic limit $\tilde {\bm{\Sigma}}_j$ has the same distribution as $\tilde {\bm{\Sigma}}_j^{(t)}$, as can be seen by comparing
Eqs.~(\ref{eq:cavEquationsMain}) and (\ref{eq:CiFinal}). Averaging
Eq.~(\ref{eq:finalSpectralCavity}) over the ensemble of matrices yields
\begin{equation}
  \begin{split}
\rho_{\bm F}(z)= -
\frac{i}{ \pi}
\partial_z^*\int d\tilde {\bm{\Sigma}} \,\tilde \omega(\tilde {\bm{\Sigma}})\,\tilde {{\Sigma}}_{21},
  \end{split}
\end{equation}
which coincides with the result obtained via the replica method.

In conclusion, upon averaging over the ensemble, the cavity equations
produce the same densities as the replica method, providing an alternative route to the replica equations.



\printbibliography

@article{Wishart1928,
  author          = {Wishart, John},
  title           = {The Generalised Product Moment Distribution in Samples from a Normal Multivariate Population},
  year            = 1928,
  volume          = {20A},
  number          = {1/2},
  month           = {Jul},
  pages           = {32},
  journal         = "Biometrika",
  url={https://www.jstor.org/stable/2331939},
  doi={https://doi.org/10.2307/2331939},
}

@article{Akemann_2008,
  author          = {Akemann, Gernot and Vivo, Pierpaolo},
  title           = {Power law deformation of {W}ishart–{L}aguerre ensembles of random matrices},
  journal         = "J. Stat. Mech. Theory Exp.",
  year            = 2008,
  volume          = 2008,
  number          = 09,
  month           = sep,
  pages           = {P09002},
  issn            = {1742-5468},
  doi             = {10.1088/1742-5468/2008/09/p09002},
  url             = {http://dx.doi.org/10.1088/1742-5468/2008/09/P09002},
  publisher       = {IOP Publishing}
}

@article{Marcenko1967,
  author = { Mar\v{c}enko, V A and Pastur, L A },
  title = { Distribution of eigenvalues for some sets of random matrices },
  journal = "Math. USSR Sb.",
  year = { 1967 },
  volume = { 1 },
  issue = { 4 },
  pages = { 457--483 },
  doi = {10.1070/SM1967v001n04ABEH001994},
  url = {https://doi.org/10.1070/SM1967v001n04ABEH001994},
}

@article{Majumdar_2012,
  author          = {Majumdar, Satya N. and Vivo, Pierpaolo},
  title           = {Number of Relevant Directions in Principal Component Analysis and {W}ishart Random Matrices},
  journal         = "Phys. Rev. Lett.",
  pages = {200601},
  year            = 2012,
  volume          = 108,
  number          = 20,
  month           = may,
  issn            = {1079-7114},
  doi             = {10.1103/physrevlett.108.200601},
  url             = {http://dx.doi.org/10.1103/PhysRevLett.108.200601},
  publisher       = {American Physical Society (APS)}
}

@article{Majumdar_2009,
  author          = {Majumdar, Satya N. and Vergassola, Massimo},
  title           = {Large Deviations of the Maximum Eigenvalue for {W}ishart and {G}aussian Random Matrices},
  journal         = "Phys. Rev. Lett.",
  pages = {060601},
  year            = 2009,
  volume          = 102,
  number          = 6,
  month           = feb,
  issn            = {1079-7114},
  doi             = {10.1103/physrevlett.102.060601},
  url             = {http://dx.doi.org/10.1103/PhysRevLett.102.060601},
  publisher       = {American Physical Society (APS)}
}

@book{Forrester_2010,
  author          = {Forrester, Peter J.},
  title           = {Log-Gases and Random Matrices },
  year            = 2010,
  publisher       = {Princeton University Press},
  isbn            = 9781400835416,
  doi             = {10.1515/9781400835416},
  url             = {http://dx.doi.org/10.1515/9781400835416},
  month           = dec
}

@article{Rogers2008,
  author          = {Rogers, Tim and Pérez Castillo, Isaac and Kühn, Reimer and
                  Takeda, Koujin},
  title           = {Cavity approach to the spectral density of sparse symmetric
                  random matrices},
  year            = 2008,
  volume          = 78,
  number          = 3,
  month           = {Sep},
  issn            = {1550-2376}, 
  journal         = "Phys. Rev. E",
    url = {https://link.aps.org/doi/10.1103/PhysRevE.78.031116},
  publisher       = {American Physical Society (APS)}
}

@article{Rogers2009,
  author          = {Rogers, Tim and Pérez Castillo, Isaac},
  title           = {Cavity approach to the spectral density of non-{H}ermitian sparse matrices},
  year            = 2009,
  volume          = 79,
  number          = 1,
  month           = {Jan},
  issn            = {1550-2376},
   journal         = "Phy. Rev. E",
    url = {https://link.aps.org/doi/10.1103/PhysRevE.79.012101},
   publisher       = {American Physical Society (APS)}
}

@article{Bouchaud2020,
  year = {2020},
  month = dec,
  publisher = {American Physical Society ({APS})},
  volume = {102},
  number = {6},
  author = {Jean-Philippe Bouchaud and Marc Potters},
  title = {Generalization of the {M}ar{\v{c}}enko-{P}astur problem},
  journal = "Phys. Rev. E",
  pages = {062117},
   url = {https://link.aps.org/doi/10.1103/PhysRevE.102.062117},
  doi={10.1103/PhysRevE.102.062117},
}

@article{PerezCastillo_2018,
  author          = {Castillo, Isaac Pérez and Metz, Fernando L.},
  title           = {Large-deviation theory for diluted {W}ishart random matrices},
  journal         = "Phys. Rev. E",
  year            = 2018,
  volume          = 97,
  number          = 3,
  month           = mar,
  issn            = {2470-0053},
  pages = {032124},
  doi             = {10.1103/physreve.97.032124},
  url             = {http://dx.doi.org/10.1103/PhysRevE.97.032124},
  publisher       = {American Physical Society (APS)}
}

@article{Laloux_1999,
  author          = {Laloux, Laurent and Cizeau, Pierre and Bouchaud, Jean-Philippe and Potters, Marc},
  title           = {Noise Dressing of Financial Correlation Matrices},
  journal         = "Phys. Rev. Lett.",
  year            = 1999,
  volume          = 83,
  number          = 7,
  month           = aug,
  pages           = {1467–1470},
  issn            = {1079-7114},
  doi             = {10.1103/physrevlett.83.1467},
  url             = {http://dx.doi.org/10.1103/PhysRevLett.83.1467},
  publisher       = {American Physical Society (APS)}
}

@article{Burda_2005,
  author          = {Burda, Zdzisław and Jurkiewicz, Jerzy and Wacław, Bartłomiej},
  title           = {Spectral moments of correlated {W}ishart matrices},
  journal         = "Phys. Rev. E",
  year            = 2005,
  pages = {026111},
  volume          = 71,
  number          = 2,
  month           = feb,
  issn            = {1550-2376},
  doi             = {10.1103/physreve.71.026111},
  url             = {http://dx.doi.org/10.1103/PhysRevE.71.026111},
  publisher       = {American Physical Society (APS)}
}

@article{Majumdar_2008,
  author          = {Majumdar, Satya N. and Bohigas, Oriol and Lakshminarayan, Arul},
  title           = {Exact Minimum Eigenvalue Distribution of an Entangled Random Pure State},
  journal         = "J. Stat. Phys.",
  year            = 2008,
  volume          = 131,
  number          = 1,
  month           = feb,
  pages           = {33–49},
  issn            = {1572-9613},
  doi             = {10.1007/s10955-008-9491-5},
  url             = {http://dx.doi.org/10.1007/s10955-008-9491-5},
  publisher       = {Springer Science and Business Media LLC}
}

@article{Telatar_1999,
  author          = {Telatar, Emre},
  title           = {Capacity of Multi‐antenna {G}aussian Channels},
  journal         = "Eur. Trans. Telecommun.",
  year            = 1999,
  volume          = 10,
  number          = 6,
  month           = nov,
  pages           = {585–595},
  issn            = {1541-8251},
  doi             = {10.1002/ett.4460100604},
  url             = {http://dx.doi.org/10.1002/ett.4460100604},
  publisher       = {Wiley}
}

@article{Verbaarschot_2000,
  author          = {Verbaarschot, J.J.M. and Wettig, T.},
  title           = {Random Matrix Theory and Chiral Symmetry in {QCD}},
  journal         = "Annu. Rev. Nucl. Part. Sci.",
  year            = 2000,
  volume          = 50,
  number          = 1,
  month           = dec,
  pages           = {343–410},
  issn            = {1545-4134},
  doi             = {10.1146/annurev.nucl.50.1.343},
  url             = {http://dx.doi.org/10.1146/annurev.nucl.50.1.343},
  publisher       = {Annual Reviews}
}

@article{Demasure_2003,
  author          = {Demasure, Yves and Janik, Romuald A.},
  title           = {Effective matter superpotentials from {W}ishart random matrices},
  journal         = "Phys. Lett. B",
  year            = 2003,
  volume          = 553,
  number          = {1–2},
  month           = jan,
  pages           = {105–108},
  issn            = {0370-2693},
  doi             = {10.1016/s0370-2693(02)03189-1},
  url             = {http://dx.doi.org/10.1016/S0370-2693(02)03189-1},
  publisher       = {Elsevier BV}
}

@article{Mezard1986,
  title={{SK} Model: The Replica Solution without Replicas},
  author={M{\'e}zard, M and Parisi, G and Virasoro, MA},
  journal= "Eur. Lett",
  volume={1},
  number={2},
  pages={77--82},
  year={1986},
doi = {10.1209/0295-5075/1/2/006},
url = {https://doi.org/10.1209/0295-5075/1/2/006},
}

@article{Peierls1936,
  title = { Statistical theory of superlattices with unequal concentrations of the components },
  author={Peierls, Rudolf},
  journal = "Proc. R. Soc. Lond. A",
  year = { 1936 },
  volume = { 154 },
  issue = { 881 },
  pages = { 207--222 },
  doi = {10.1098/rspa.1936.0047},
  url = {http://doi.org/10.1098/rspa.1936.0047},
}

@article{Mezard2001,
  author = { M{\'e}zard, M. and Parisi, G. },
  title = { The {B}ethe lattice spin glass revisited },
  journal = "Eur. Phys. J. B",
  year = { 2001 },
  volume = { 20 },
  issue = { 2 },
  pages = { 217--233 },
  doi = {10.1007/PL00011099},
  url = {http://doi.org/10.1007/PL00011099},
}

@article{RamosSanchez2020,
  author          = {Pérez Castillo, Isaac  and Guzmán-González, Edgar and
                  Sánchez, Antonio Tonatiúh Ramos and Metz, Fernando L.},
  title           = {Analytic approach for the number statistics of
                  non-{H}ermitian random matrices},
  year            = 2021,
  volume          = 103,
  pages = {062108},
  number          = 6,
  month           = {Jun},
  issn            = {2470-0053},
  doi             = {10.1103/physreve.103.062108},
  url             = {http://dx.doi.org/10.1103/PhysRevE.103.062108},
  journal         = "Phys. Rev. E",
  publisher       = {American Physical Society (APS)}
}

@article{Bouchaud2006,
  author          = {Bouchaud, J.-P. and Laloux, L. and Miceli, M. A. and Potters, M.},
  title           = {Large dimension forecasting models and random singular value spectra},
  year            = 2006,
  volume          = 55,
  number          = 2,
  month           = {May},
  pages           = {201–207},
  issn            = {1434-6036},
  doi             = {10.1140/epjb/e2006-00204-0},
  url             = {http://dx.doi.org/10.1140/epjb/e2006-00204-0},
  journal         = "Eur. Phys. J. B",
  publisher       = {Springer Science and Business Media LLC}
}

@article{Bishop2018,
  author          = {Bishop, Adrian N. and Del Moral, Pierre and Niclas, Angèle},
  title           = {An Introduction to {W}ishart Matrix Moments},
  year            = 2018,
  volume          = 11,
  number          = 2,
  pages           = {97–218},
  issn            = {1935-8245},
  doi             = {10.1561/2200000072},
  url             = {http://dx.doi.org/10.1561/2200000072},
  journal         = "Found. Trends Mach. Learn.",
  publisher       = {Now Publishers}
}

@book{Mehta2004,
  title           = {Random matrices},
  author          = {Mehta, Madan Lal},
  year            = {2004},
  isbn            = {9780120884094},
  publisher       = {Elsevier},
  address         = {Amsterdam Heidelberg},
  edition         = {3},
  number          = {142},
  series          = "Pure Appl. Math.",
}

@book{Akemann2015,
  title           = {The Oxford Handbook of Random Matrix Theory},
  author          = {Akemann, Gernot and Baik, Jinho and Di Francesco, Philippe},
  year            = {2015},
  month           = {Sep},
  doi             = {10.1093/oxfordhb/9780198744191.001.0001},
  url             = {http://dx.doi.org/10.1093/oxfordhb/9780198744191.001.0001},
  isbn            = {9780191873997},
  publisher       = {Oxford University Press},
  address         = {New York}
}

@article{Tulino2004,
  author          = {Tulino, Antonia M. and Verdú, Sergio},
  title           = {Random Matrix Theory and Wireless Communications},
  year            = 2004,
  volume          = 1,
  number          = 1,
  pages           = {1–182},
  issn            = {1567-2328},
  doi             = {10.1561/0100000001},
  url             = {http://dx.doi.org/10.1561/0100000001},
  journal         = "Found. Trends Commun. Inf. Theory",
  publisher       = {Now Publishers}
}

@article{Edwards1976,
  author          = {Edwards, S F and Jones, R C},
  title           = {The eigenvalue spectrum of a large symmetric random matrix},
  year            = 1976,
  volume          = 9,
  number          = 10,
  month           = {Oct},
  pages           = {1595–1603},
  issn            = {1361-6447},
  doi             = {10.1088/0305-4470/9/10/011},
  url             = {http://dx.doi.org/10.1088/0305-4470/9/10/011},
  journal         = "J. Phys. A",
  publisher       = {IOP Publishing}
}

@article{Erdos1959,
  author          = {Erd\H{o}s, P. and R\'{e}nyi, A.},
  journal         = "Publ. Math. Debrecen",
  title           = {On Random Graphs {I}},
  volume          = {6},
  pages           = {290--297},
  year            = {1959}
}

@book{Mezard1986_2,
  title = {Spin Glass Theory and Beyond: An Introduction to the Replica Method and Its Applications},
  ISBN = {9789812799371},
  ISSN = {1793-1436},
  url = {http://dx.doi.org/10.1142/0271},
  DOI = {10.1142/0271},
  journal = "World Sci. Lect. Notes Phys.",
  publisher = {World Scientific},
  author = {Mezard,  M and Parisi,  G and Virasoro,  M},
  year = {1986},
  month = nov
}

@article{PerezCastillo2022,
  author          = {Pérez Castillo, Isaac},
  title           = {Spectral properties of the generalized diluted {W}ishart ensemble},
  journal         = "J. Physics: Complex.",
  year            = 2022,
  volume          = 3,
  number          = 4,
  month           = oct,
  pages           = 045001,
  issn            = {2632-072X},
  doi             = {10.1088/2632-072x/ac956d},
  url             = {http://dx.doi.org/10.1088/2632-072X/ac956d},
  publisher       = {IOP Publishing}
}

@article{Akemann2011,
  author          = {Akemann, G.},
  title           = {Non-{H}ermitian Extensions of {W}ishart Random Matrix Ensembles},
  journal         = "Acta Phys. Pol. B",
  year            = 2011,
  volume          = 42,
  number          = 5,
  pages           = 901,
  issn            = {1509-5770},
  doi             = {10.5506/aphyspolb.42.901},
  url             = {http://dx.doi.org/10.5506/APhysPolB.42.901},
  publisher       = {Jagiellonian University}
}

@article{Ginibre_1965,
  author          = {Ginibre, Jean},
  title           = {Statistical Ensembles of Complex, Quaternion, and Real Matrices},
  journal         = "J. Math. Phys.",
  year            = 1965,
  volume          = 6,
  number          = 3,
  month           = mar,
  pages           = {440–449},
  issn            = {1089-7658},
  doi             = {10.1063/1.1704292},
  url             = {http://dx.doi.org/10.1063/1.1704292},
  publisher       = {AIP Publishing}
}

@article{Metz_2016,
  author          = {Metz, Fernando L. and Pérez Castillo, Isaac},
  title           = {Large Deviation Function for the Number of Eigenvalues of Sparse Random Graphs Inside an Interval},
  journal         = "Phys. Rev. Lett.",
  year            = 2016,
  volume          = 117,
  number          = 10,
  pages = {104101},
  month           = sep,
  issn            = {1079-7114},
  doi             = {10.1103/physrevlett.117.104101},
  url             = {http://dx.doi.org/10.1103/PhysRevLett.117.104101},
  publisher       = {American Physical Society (APS)}
}

@book{Nishimori_2001,
  title     = {Statistical Physics of Spin Glasses and Information Processing: An Introduction},
  author    = {Hidetoshi Nishimori},
  year      = {2001},
  address   = {Oxford; New York},
  publisher = {Oxford University Press},
  series    = "Int. Monogr. Phys.",
  doi       = "10.1093/acprof:oso/9780198509417.001.0001",
  volume    = {111},
  isbn      = {978-0-19-850940-0}
}

@online{HazarikaPaul2024,
  author    = {Javed Hazarika and Debashis Paul},
  title     = {Limiting Spectral Distribution of a Random Commutator Matrix},
  year      = {2024},
  note      = {Preprint, arXiv:2409.16780},
  url       = {https://arxiv.org/abs/2409.16780},
  doi       = {10.48550/arXiv.2409.16780}
}

@article{Kuhn_2008,
  author          = {Kühn, Reimer},
  title           = {Spectra of sparse random matrices},
  journal         = "J. Phys. A: Math. Theor.",
  year            = 2008,
  volume          = 41,
  number          = 29,
  month           = jun,
  pages           = 295002,
  issn            = {1751-8121},
  doi             = {10.1088/1751-8113/41/29/295002},
  url             = {http://dx.doi.org/10.1088/1751-8113/41/29/295002},
  publisher       = {IOP Publishing}
}

@article{Rotter_2009,
  author          = {Rotter, Ingrid},
  title           = {A non-{H}ermitian {H}amilton operator and the physics of open quantum systems},
  journal         = "J. Phys. A: Math. Theor.",
  year            = 2009,
  volume          = 42,
  number          = 15,
  month           = mar,
  pages           = 153001,
  issn            = {1751-8121},
  doi             = {10.1088/1751-8113/42/15/153001},
  url             = {http://dx.doi.org/10.1088/1751-8113/42/15/153001},
  publisher       = {IOP Publishing}
}

@book{Moiseyev_2011,
  author          = {Moiseyev, Nimrod},
  title           = {Non-{H}ermitian Quantum Mechanics},
  year            = 2011,
  publisher       = {Cambridge University Press},
  isbn            = 9780511976186,
  doi             = {10.1017/cbo9780511976186},
  url             = {http://dx.doi.org/10.1017/CBO9780511976186},
  month           = feb
}

@article{Allesina_2015,
  author          = {Allesina, Stefano and Tang, Si},
  title           = {The stability–complexity relationship at age 40: a random matrix perspective},
  journal         = "Popul. Ecol.",
  year            = 2015,
  volume          = 57,
  number          = 1,
  month           = jan,
  pages           = {63–75},
  issn            = {1438-390X},
  doi             = {10.1007/s10144-014-0471-0},
  url             = {http://dx.doi.org/10.1007/s10144-014-0471-0},
  publisher       = {Wiley}
}

@article{Burda_2004,
  author          = {Burda, Zdzisław and Jurkiewicz, Jerzy and Nowak, Maciej A. and Papp, Gabor and Zahed, Ismail},
  title           = {Free {L}évy matrices and financial correlations},
  journal         = "Phys. A: Stat. Mech. Its Appl.",
  year            = 2004,
  volume          = 343,
  month           = nov,
  pages           = {694–700},
  issn            = {0378-4371},
  doi             = {10.1016/j.physa.2004.05.049},
  url             = {http://dx.doi.org/10.1016/j.physa.2004.05.049},
  publisher       = {Elsevier BV}
}

@article{Hatano_1996,
  author          = {Hatano, Naomichi and Nelson, David R.},
  title           = {Localization Transitions in Non-{H}ermitian Quantum Mechanics},
  journal         = "Phys. Rev. Lett.",
  year            = 1996,
  volume          = 77,
  number          = 3,
  month           = jul,
  pages           = {570–573},
  issn            = {1079-7114},
  doi             = {10.1103/physrevlett.77.570},
  url             = {http://dx.doi.org/10.1103/PhysRevLett.77.570},
  publisher       = {American Physical Society (APS)}
}

@article{Shnerb_1998,
  author          = {Shnerb, Nadav M. and Nelson, David R.},
  title           = {Winding Numbers, Complex Currents, and Non-{H}ermitian Localization},
  journal         = "Phys. Rev. Lett.",
  year            = 1998,
  volume          = 80,
  number          = 23,
  month           = jun,
  pages           = {5172–5175},
  issn            = {1079-7114},
  doi             = {10.1103/physrevlett.80.5172},
  url             = {http://dx.doi.org/10.1103/PhysRevLett.80.5172},
  publisher       = {American Physical Society (APS)}
}

@article{Metz_2019,
  author          = {Lucas Metz, Fernando and Neri, Izaak and Rogers, Tim},
  title           = {Spectral theory of sparse non-{H}ermitian random matrices},
  journal         = "J. Phys. A: Math. Theor.",
  year            = 2019,
  volume          = 52,
  number          = 43,
  month           = oct,
  pages           = 434003,
  issn            = {1751-8121},
  doi             = {10.1088/1751-8121/ab1ce0},
  url             = {http://dx.doi.org/10.1088/1751-8121/ab1ce0},
  publisher       = {IOP Publishing}
}

@article{Neri_2016,
  author          = {Neri, Izaak and Metz, Fernando Lucas},
  title           = {Eigenvalue Outliers of Non-{H}ermitian Random Matrices with a Local Tree Structure},
  journal         = "Phys. Rev. Lett.",
  pages = {030602},
  year            = 2016,
  volume          = 117,
  number          = 22,
  month           = nov,
  issn            = {1079-7114},
  doi             = {10.1103/physrevlett.117.224101},
  url             = {http://dx.doi.org/10.1103/PhysRevLett.117.224101},
  publisher       = {American Physical Society (APS)}
}

@article{Neri_2012,
  author          = {Neri, I. and Metz, F. L.},
  title           = {Spectra of Sparse Non-{H}ermitian Random Matrices: An Analytical Solution},
  pages = {030602},
  journal         = "Phys. Rev. Lett.",
  year            = 2012,
  volume          = 109,
  number          = 3,
  month           = jul,
  issn            = {1079-7114},
  doi             = {10.1103/physrevlett.109.030602},
  url             = {http://dx.doi.org/10.1103/PhysRevLett.109.030602},
  publisher       = {American Physical Society (APS)}
}

@article{Nagao_2007,
  author          = {Nagao, Taro and Tanaka, Toshiyuki},
  title           = {Spectral density of sparse sample covariance matrices},
  journal         = "J. Phys. A: Math. Theor.",
  year            = 2007,
  volume          = 40,
  number          = 19,
  month           = apr,
  pages           = {4973–4987},
  issn            = {1751-8121},
  doi             = {10.1088/1751-8113/40/19/003},
  url             = {http://dx.doi.org/10.1088/1751-8113/40/19/003},
  publisher       = {IOP Publishing}
}
\end{document}